\documentclass[superscriptaddress,altaffilletter,amsmath,amssymb,aps,preprint,floatfix]{revtex4-1}
\pdfoutput=1 
\usepackage[english]{babel}
\usepackage{multirow}
\usepackage{hhline}
\usepackage{bigdelim}
\usepackage{enumerate}
\usepackage{relsize}
\usepackage{float}
\usepackage{textcomp}
\usepackage{hyperref}
\usepackage[compact]{titlesec}
\titlespacing{\section}{0pt}{3ex}{2ex}
\titlespacing{\subsection}{0pt}{2ex}{2ex}
\titlespacing{\subsubsection}{0pt}{1.5ex}{1.5ex}

\usepackage{longtable}
\usepackage[dvipsnames]{xcolor}
\usepackage{graphicx}
\usepackage{color}
\usepackage{dcolumn}
\usepackage{bm}
\usepackage{hyperref}
\usepackage[mathlines]{lineno}
\usepackage{threeparttable} 
\usepackage{booktabs} 
\usepackage{siunitx}
\usepackage{placeins}
\usepackage{xspace}
\usepackage{tabularx}
\usepackage{tablefootnote}
\usepackage{adjustbox}
\newcommand{\rn}[2]{$^{#1}$#2}
\newcommand{\mysum}[3]{\sum\limits_{#1}^{#2}{#3}} 
\newcommand{\myprod}[3]{\prod\limits_{#1}^{#2}{#3}} 

\begin{document}
	
\preprint{APS/123-QED}
	
\title{Influence of atomic modeling on electron capture and shaking processes}
	
\author{A. Andoche}
\author{L. Mouawad}
\author{P.-A. Hervieux}
\email{hervieux@unistra.fr}
\affiliation{Universit{\'e} de Strasbourg, CNRS, Institut de Physique et Chimie des Mat{\'e}riaux de Strasbourg, UMR 7504, F-67000 Strasbourg, France
}%
\author{X. Mougeot}
\email{xavier.mougeot@cea.fr}
\affiliation{Universit{\'e} Paris-Saclay, CEA, List, Laboratoire National Henri Becquerel (LNE-LNHB), F-91120 Palaiseau, France
}%
\author{J. Machado}
\email{jfd.machado@fct.unl.pt}
\author{J.P. Santos}
\affiliation{Laboratory of Instrumentation, Biomedical Engineering and Radiation Physics (LIBPhys-UNL), Department of Physics, NOVA School of Science and Technology, NOVA University Lisbon, 2829-516 Caparica, Portugal
}%
 
\date{January 23, 2024}

\begin{abstract}
Ongoing experimental efforts to measure with unprecedented precision electron-capture probabilities challenges the current theoretical models. The short range of the weak interaction necessitates an accurate description of the atomic structure down to the nucleus region. A recent electron-capture modeling has been modified in order to test the influence of three different atomic descriptions on the decay and shaking probabilities. To this end, a specific atomic modeling was developed in the framework of the relativistic density-functional theory, exploring several exchange-correlation functionals and self-interaction-corrected models. It was found that the probabilities of total shaking, tested on both photoionization and electron-capture processes, depend strongly on the accuracy of the atomic modeling. Predictions of capture probabilities have been compared with experimental values evaluated from available published data for different radionuclides from \rn{7}{Be} to \rn{138}{La}. New high-precision measurements are strongly encouraged because the accuracy of the current experimental values is insufficient to test the models beyond the inner shells.
\end{abstract}

\maketitle

\section{Introduction}

Electron capture is a low-energy weak interaction, wherein a proton embedded in a nucleus, absorbs an atomic electron. This results in the formation of a nucleus with a reduced atomic number, accompanied by the emission of a neutrino with a precisely defined energy. Upon undergoing the transition, if the daughter nucleus is not in  its ground state, its deexcitation occurs through a cascade of $\gamma$ transitions. This sequence ultimately results in the emission of either $\gamma$ rays or internal-conversion electrons, the latter inducing the creation of atomic vacancies. Furthermore, immediately following the transition, the daughter atom is neutrally charged but in an excited state, characterized by a neutral atom with a vacancy. The subsequent relaxation of this state results in the emission of x rays and Auger electrons throughout the propagation of the vacancies until their disappearance.

 Ground-state to ground-state transitions are very difficult to measure, especially for low-mass nuclei, because only x rays and Auger electrons of very low energies can be detected. Standardization of such pure electron-capture emitters in radionuclide metrology is challenging and directly depends on the knowledge of the capture probabilities, which most often come from theory. Recently, a new international reference system, called ESIR \cite{Coulon20}, has been established by the International Bureau of Weights and Measures (BIPM) based on activity measurements carried out with the liquid scintillation counting (LSC) technique. The computed LSC efficiency strongly decreases at very low energy and a precise modeling of the atomic rearrangement is necessary, which depends on the number of initial vacancies \cite{Broda07}.
	
What is problematic for radionuclide metrology can become an advantage in nuclear medicine for cancer treatment when the radioactive nuclei are correctly vectorized into the human body. The vast majority of Auger electrons are emitted with a few keV kinetic energy, deposited in a range from nanometers to micrometers. This property makes them highly promising for an accurate, well-controlled internal radiotherapy at the cell or even the DNA level. Many short-lived nuclei decaying by electron capture have been studied in this perspective \cite{Bave18,Ku19}.
	
Long-lived isotopes can also be useful in other contexts. The decommissioning of legacy nuclear sites necessitates the radioactivity inventory of various materials. When nondestructive techniques such as $\gamma$ spectrometry are impossible, precise measurements of samples are used together with scaling methods to estimate the total amount of radioactivity \cite{Hou05,Lee21}. Some long-lived pure electron-capture emitters have been used to investigate the age of the Solar System \cite{Huss09,Jorg12} and the irradiation history of meteorites and lunar samples \cite{Kubik86}.
	
Every application refers to evaluated nuclear decay data, which can be taken either in the ENSDF (Evaluated Nuclear Structure Data File) database \cite{ENSDF92} or in the DDEP (Decay Data Evaluation Project) database \cite{DDEP17}, the latter being recommended by the BIPM for metrology purposes. In ENSDF evaluations, the properties of $\beta$ transitions and electron captures have been calculated for the last 50 years with the LogFT code \cite{Gove71}, which is based on an approximate theoretical model limited by the computing power available in the 1970s. Over the past decade, the BetaShape code has been developed by one of the authors in order to improve the theoretical predictions for $\beta$ transitions and electron captures \cite{Mougeot15,Mougeot19}. Its modeling takes into account several additional physical phenomena and provides much more detailed information required by different communities. The DDEP Collaboration has already adopted this new code for its nuclear data evaluations. Recently, the International Network of Nuclear Structure and Decay Data Evaluators (NSDD) has also adopted it for the future ENSDF evaluations.
	
Capture probabilities from BetaShape have been compared with a selection of measurements available in the literature, concluding to the need of new high-precision measurements to validate and constrain the theoretical models \cite{Mougeot18}. This played a crucial role in the inception of the European metrology project MetroMMC \cite{MetroMMC20}, which was dedicated to advancing our comprehension of electron-capture decay and the subsequent processes involved in atomic relaxation. The ongoing European metrology project PrimA-LTD \cite{PrimaLTD} also addresses this topic, one of its ambitions being the measurement of the \rn{55}{Fe} capture spectrum with unprecedented precision. Such high-precision measurements challenge the theoretical predictions, for which the accuracy of the atomic modeling is essential. Indeed, as the electron-capture process takes place inside the nucleus, the description of the electronic properties of atoms must be as precise as possible, in particular in this region of space. In addition, one can also wonder about the role played by electron correlations in the decay process. Trying to answer these two issues constitute the main goal of this work.

To this end, we have studied the influence of atomic modeling on electron capture and shaking processes considering three different approaches. In Sec. \ref{sec:theo_meth}, we first present the framework used to calculate these two processes, as implemented in the BetaShape code. Next, we summarize two existing atomic models: the basic BetaShape model and a much more accurate one, the Multiconfiguration Dirac-Fock (MCDF) method. Finally, we describe in detail a model specifically developed for this work (called KLI for Krieger, Li and Iafrate) and built within the framework of relativistic density-functional theory with local density approximation (RLDA). In Sec. \ref{sec:gen_res}, we compare our theoretical predictions for binding energies, shaking and capture probabilities between each other, and to experimental values when available. 

For ease of comparison with the literature, we use in the theoretical descriptions the usual notation for atomic orbitals, i.e., 1$s_{1/2}$ or 2$p_{3/2}$, and Siegbahn's notation when comparing with experimental results, i.e., $K$ or $L_{3}$, respectively. The probability of decaying through the capture of an electron in a $K$, $L$, etc. shell is denoted $P_K$, $P_L$, etc., respectively.
	
\section{Theoretical methods}
\label{sec:theo_meth}
 
Our modeling of electron-capture decay, which includes the shaking process, is based on the formalism from Refs. \cite{Bamb77,Behrens82} and has already been described in detail in Refs. \cite{Mougeot18,Mougeot19}. In this section, we focus on the dependency of this modeling on the atomic wave functions, and on the determination of the wave functions themselves.
	
\subsection{Electron capture}
\label{sec:elec_capt}
	
The low rest mass of the $\beta$ particle compared with the transition energies necessitates a fully relativistic formalism of electron-capture decay, usually expressed in relativistic units ($\hbar = m_e = c = 1$). Considering spherical symmetry, radial and angular parts of the wave functions can be separated. An atomic orbital is then characterized by its quantum numbers ($n$,$\kappa$), the latter being the eigenvalue of the operator $K = \beta(\vec{\sigma}\cdot\vec{L}+1)$ defined from the Dirac ($\beta$) and Pauli ($\vec{\sigma}$) matrices and the angular-momentum operator ($\vec{L}$). 
 
Atomic levels being degenerate, one has to introduce their relative occupation number $n_{n\kappa}$. In addition, each wave function is characterized by its Coulomb amplitude $\beta_{n\kappa}$, as defined in Refs. \cite{Bamb77,Behrens82}. This quantity has to be computed and is simply the value of the wave function at the origin ($r=0$) for $\kappa=-1$, i.e., in the case of $s_{1/2}$ orbitals.
	
The transition probability per unit of time can be derived by applying first-order time-dependent perturbation theory and results in the sum of the capture probability of each subshell:
\begin{eqnarray}
\label{eq:lambda_ec}
\lambda_{\varepsilon} &=& \dfrac{G^{2}_{\beta}}{2 \pi^3} \mysum{n\kappa}{}{ \frac{\pi}{2} n_{n\kappa} C_{n\kappa} q^{2}_{n\kappa} \beta^{2}_{n\kappa} B_{n\kappa} \left( 1 + \sum_{n'\kappa'}{} P_{n\kappa}^{n'\kappa'} \right) } \nonumber \\
&=& \mysum{n\kappa}{}{P_{n\kappa}} \text{,}
\end{eqnarray}
\noindent with $G_{\beta}$ being the Fermi coupling constant and $q_{n\kappa}$ the neutrino momentum. In the present work, we do not consider any electron-capture decay for which the available energy would be sufficient for a competition with a $\beta^{+}$ transition.
	
The quantity $C_{n\kappa}$ couples the lepton and nucleon wave functions. For allowed and forbidden unique transitions, the nuclear dependency acts as a constant factor. Therefore, determining ratios of relative capture probabilities prevents from introducing any nuclear structure. In such cases,
\begin{equation}
 \label{eq:C_nk}
C_{n\kappa} \propto \dfrac{p_{n\kappa}^{2(k-1)}~q_{n\kappa}^{2(L-k)}}{(2k-1)![2(L-k)+1]!} \text{,}
\end{equation}
with $k = |\kappa|$, $p_{n\kappa}$ the electron momentum and $L = \left| J_i - J_f \right|$ the difference between the total angular momenta of initial and final nuclear levels. 
	
The quantity $B_{n\kappa}$ is a correction that accounts for overlap and exchange effects. The former is induced by the capture process that changes the nucleus charge, leading to an imperfect overlap of the initial and final wave functions of the spectator electrons. The latter arises from electrons being identical particles. The two following methods, available in the literature, have been considered in the BetaShape code for calculating $B_{n\kappa}$. Bahcall's approach \cite{Bahcall65} only considers the first three $s_{1/2}$ orbitals, assumes a complete set of states for the other orbitals and makes use of the closure property to sum over the continuum states. Vatai's approach \cite{Vatai70} considers in addition the exchange term of the $4s_{1/2}$ orbital and the overlap correction of each subshell, but does not use the closure property. The generalization of theses approaches \cite{Mougeot18} leads to $B_{n \kappa} = \left| b_{n \kappa} / \beta_{n \kappa} \right|^2$ with 
	\begin{equation}
	b_{n\kappa} = t_{n\kappa} \left[ \myprod{m \kappa \neq n \kappa}{}{\langle (m \kappa)'|(m \kappa)\rangle} \right] \times \left[ \beta_{n\kappa} - \mysum{m \kappa \neq n \kappa}{}{ \beta_{m\kappa} \dfrac{\langle (m \kappa)'|(n \kappa)\rangle}{\langle(m \kappa)'|(m \kappa)\rangle} } \right] \text{,}
	\end{equation}
	where $\langle(m \kappa)'|(m \kappa)\rangle$ is the overlap of the atomic wave functions between the initial $(m \kappa)$ state and the final $(m \kappa)'$ state, and correspondingly for the other overlaps. Bahcall's model is given by $t_{n\kappa} = 1$ and Vatai's model by
	\begin{equation}
	t_{n\kappa} = \langle(n \kappa)'|(n \kappa)\rangle^{n_{n\kappa} - 1/(2|\kappa|)} \times \left[ \myprod{m \kappa \neq n \kappa}{}{\langle(m \kappa)'|(m \kappa)\rangle^{n_{m\kappa} - 1} } \right] \times \left[ \myprod{\tiny \begin{array}{c} m \mu \\ \mu \neq \kappa \\ \end{array} \normalsize}{}{ \langle(m \mu)'|(m \mu)\rangle^{n_{m\mu}} } \right]\text{.}
	\end{equation}
The difference between the Bahcall and Vatai models in the resulting capture probabilities is used in BetaShape as a theoretical uncertainty component. The other component comes from input data and corresponds to the propagation of the uncertainty on the transition energy. The factor in parentheses in Eq. (\ref{eq:lambda_ec}) corrects for the shaking effects, i.e., internal excitation (shake-up) and internal ionization (shake-off). These effects take into account additional final states in which other atomic vacancies are present with the initial vacancy due to capture. The modeling implemented in BetaShape is limited to a single additional vacancy due to shaking. 
The separate calculation of each process requires the knowledge of all unoccupied bound states up to the Fermi level for shake-up, and summing over the infinite continuum states for shake-off. However, the determination of capture probabilities only requires the total probability $P_{n \kappa}^{n'\kappa'}$ of creating a secondary vacancy in an orbital $(n' \kappa')$ consecutive to the capture of an electron in an orbital $(n \kappa)$. For better readability, we denote in the following $(n' \kappa')$ and $(n \kappa)$ by $(i)$ and $(j)$, respectively. Such a probability is simpler to calculate when considering the nonshaking probability, and can be expressed as \cite{Crase79}:
\begin{equation}
P_{j}^i=1-|\langle(i_j)'|(i)\rangle|^{2N_{i}} - \frac{N_{i}}{2k_{i}} \sum  \limits_{l\ne i} N^{\prime l}_{j} |\langle(l_j)'|(i)\rangle|^2 \text{,}
\label{eq:sudden_shakes}
\end{equation}
where $N_i$ indicates the occupancy number of the parent's $(i)$ orbital and $N^{\prime l}_{j}$ denotes the occupancy number of the daughter's $(l)$ orbital after capture in the $(j)$ orbital. These effects are already taken into account in the Bahcall modeling of overlap and exchange with the use of the closure property. On the contrary, they are not included in Vatai's approach. The shaking correction is therefore only applied along with Vatai modeling. 
	
Finally, the possibility of radiative capture process is taken into account with radiative corrections, for which calculation depends on the atomic energies. However, they are only significant when a $\beta^{+}$ transition competes, i.e., for the determination of capture-to-positron ratios.

\subsection{Wave functions}
\subsubsection{BetaShape wave functions}
	
The relativistic electron wave functions in BetaShape are determined following essentially the method of Behrens and Bühring \cite{Behrens82}, as described in detail in Refs. \cite{Mougeot19,Mougeot14}. Consider a neutral atom whose orbitals are filled according to Madelung's rule. The potential created by the nucleus is modeled by that of a uniformly charged sphere, while the one produced by the bound electrons is constructed from the self-consistent Dirac-Hartree-Fock-Slater model described in Ref. \cite{Salvat87}. For bound states, an exchange potential is added that includes a constant prefactor. The latter is adjusted in order to force the convergence procedure to tabulated atomic energies. 
	
As underlined by Vatai in Ref. \cite{Vatai70}, the vacancy created in a subshell by the capture process has a significant influence on all the orbitals. This hole effect is corrected employing first-order time-independent perturbation theory \cite{Mougeot18}, as proposed by Vatai.
	
The tabulated Dirac-Fock binding energies from Desclaux \cite{Desclaux73} were used in a first study \cite{Mougeot18}. However, comparison with a selection of experimental values showed that the accuracy of the electronic wave functions was not sufficient to distinguish between Bahcall and Vatai models of overlap and exchange correction. In a second study \cite{Mougeot19}, binding energies from Kotochigova \textit{et al.} \cite{Kot97} recommended on the NIST website \cite{NIST09} were considered. They were determined using a point-charge model to describe the nucleus potential and using the relativistic local density approximation, including electronic correlation. Good agreement with the experimental data was found and, as expected, the Vatai overlap and exchange model, corrected for shaking proved to be more accurate than the Bahcall one.
	
It is clear that this method for determining the electronic wave functions is inconsistent. First, because a realistic model must use an extended charge distribution to describe the nucleus, whereas Kotochigova's approach uses a point-charge distribution. Second, because no correlation is included in the adjusted potential of BetaShape (adjustment via the constant prefactor of the exchange potential). Finally, the method used to obtain changes due to hole creation is approximate, since it is based on a perturbative approach. 
 
On the contrary, the model outlined in the present work, which is further elucidated below, demonstrates complete consistency. Depending on the atom under consideration, the electronic properties are calculated using either the MCDF method, which is the most sophisticated model to date, but is limited to light atoms with a few electronic orbitals, or the KLI model, which provides a realistic description for all atoms. A specific version of the BetaShape code has been developed to use wave functions, binding energies and electronic configuration from MCDF and KLI models. The approximate correction of the hole effect has been removed and this effect is exactly taken into account by determining the wave functions for an electronic configuration with a vacancy due to capture.

\subsubsection{Multiconfiguration Dirac-Fock (MCDF) method}
\label{sec:MCDF}
 
The MCDF wave functions were computed using the Multiconfiguration Dirac-Fock and General Matrix Element (MCDFGME) code developed by Desclaux and Indelicato~\cite{DESCLAUX1975,INDELICATO1990}. Nuclear size effects are taken into account by considering a two-parameter Fermi distribution \cite{Palffy10}. Atomic masses and nuclear radii are taken respectively from the tables given in Refs. \cite{AUDI03} and~\cite{ANGELI2013}. For a detailed description of the MCDF method, we refer the reader to~\cite{DESCLAUX1975,GRANT198837,INDELICATO1995,PARPIA1996}. 

We provide here only a brief description of the method. The effective relativistic Hamiltonian for the $N$-electron system in the MCDF method reads
\begin{equation}
H_{\text{DCB}}=\sum_{i=1}^{N}h_i^{\text{D}} + \sum_{i=1}^{N-1}\sum_{j=i+1}^{N}V_{ij}^{\text{CB}} \text{,}
\end{equation}
where $h_i^{\text{D}}$ is the Dirac one-electron Hamiltonian of the $i$-th electron. The term $V_{ij}^{\text{CB}}$ describes the Coulomb repulsion and the Breit interaction (magnetic interaction and retardation) between the $i$-th and the $j$-th electrons. In this context, the method is characterized by the optimization of the $N$-electron wave function obtained by minimizing the total energy of the system through the self-consistent field approximation. The $N$-electron many-body wave function for a state $s$ with total angular momentum $J$, its projection on the chosen direction $M$ and parity $p$, is assumed to be expressed as
\begin{equation}
    \Psi_s(JM^p)=\sum_m c_m(s)\Phi(\gamma_m JM^p)\text{,}
\end{equation}
where $\Phi(\gamma_m JM^p)$ are the Configuration State Functions (CSF) expressed in the form of Slater determinants or a linear combination of Slater determinants built from one-electron Dirac spinors. The mixing coefficients are expressed by $c_m(s)$ for state $s$ while $\gamma_m$ represents all the required information to uniquely define a given CSF.

In the MCDF method, in order to obtain high-precision results on energies and wave functions, it is necessary to include all singly and doubly excited CSFs up to a certain principal quantum number $n$, which must be greater than the principal quantum number of the valence shell. 

The effect of electronic correlations is particularly important for open-shell atomic systems or atoms with a few electrons. Due to the computational time, which becomes prohibitive for atoms with more than a dozen of electrons, we have applied this method only to $^7$Be. It is important to stress that the use of this \textit{ab initio} method makes it possible to study, with great precision, the role played by electronic correlations in the electron-capture process. This is done by comparing the shake and capture probabilities obtained with and without taking into account electronic correlations. The latter have been calculated for all singly and doubly CSFs excited up to $n=5$. We found that the change due to correlations is negligible in the shaking and capture probabilities, although the energies change with increasing configuration number, i.e., with increasing correlation (see Table \ref{tab:bind_ener}).

Another level of complexity arises from the fact that we have to consider atomic systems with a vacancy (due to the capture process) in the inner electron shell. In this case, multiple energy levels can be generated due to the coupling of angular momentum between that of the vacancy state and that of the other electrons. However, for all the atomic systems investigated in this work ($^7$Be, $^{37}$Ar, $^{41}$Ca and $^{109}$Cd, all of which are initially closed-shell atoms), we have considered only one electronic level per vacancy, which makes the calculations less time consuming.

\subsubsection{RLDA-KLI}

To describe the electron-capture process, one has to work in the framework of relativistic quantum mechanics for describing the electronic part of the atom. In this context, it is also mandatory to include correlation effects -- beyond mean-field -- and nuclear finite-size effects by using a precise description of the nucleus charge distribution -- beyond the usual point-charge approximation. Indeed, the penetration of the electron wave function inside the nucleus can be important for heavy elements. 

As previously discussed in \ref{sec:MCDF}, one of the most popular and powerful method for including correlation effects at a relativistic level is the MCDF model, which uses a linear combination of Slater determinants to approximate the electron many-body wave function \cite{DESCLAUX1975}. However as mentioned above, the use of MCDF is restricted to atoms having a relatively low atomic number (up to $Z=12$) due to the exponential increase with atomic number of the configurations involved in the computation.

An alternative approach, although \textit{a priori} less accurate, is provided by the relativistic density-functional theory (RDFT) within the local density approximation (RLDA) \cite{Dreizler1998}. The latter is much less time consuming, easier to use and able to cover a large number of radionuclides. In the present work, we have developed a specific calculation code based on this method. The main ingredients of RLDA are reminded below and we refer to \cite{Rajagopal_1978,MacDonald_1979,Engel02} for more details. In the following, spherical symmetry is assumed and atomic units are used if not otherwise stated. 

In RDFT, one must solve the one electron Dirac equation
\begin{equation}
	\left[\vec{\alpha}.\vec{p}/\alpha + \beta/\alpha^2 +V_{\mathrm{eff}}[\rho](r)\right]\Psi_i(\vec{r})=\varepsilon_i \Psi_i(\vec{r}) \text{,}
\end{equation}
where $\vec{\alpha}$ and $\beta$ are the Dirac matrices \cite{Strange08}, $\alpha$ is the fine-structure constant, $\Psi_i$ is a bi-spinor and $V_{\mathrm{eff}}[\rho]$ is an effective spherical one-particle potential. $T$ is the kinetic energy that introduces a coupling between the small and large components of the Dirac wave function and $\beta/\alpha^2$ is the electron rest energy. The electron density is given by
	\begin{equation}
		\rho(r)= \sum_{i=1}^{N_{\mathrm{orb}}} \Psi_i^{\dagger}(\vec{r})\Psi_i(\vec{r}) = \sum_{i=1}^{N_{\mathrm{orb}}}\rho_i(r) \text{,}
	\end{equation}
where $N_{\mathrm{orb}}$ is the number of orbitals.

Within the framework of RLDA, the effective potential can be expressed as
	\begin{equation}
		V_{\mathrm{eff}}[\rho](r)= V_{\mathrm{H}}[\rho](r) + V_{\mathrm{xc}}[\rho](r) + V_{\mathrm{ext}}(r) \text{,}
	\end{equation}
where $V_{\mathrm{H}}[\rho](r) = \int \frac{\rho(r')}{\left|\vec{r}-\vec{r}' \right|}d\vec{r}'$ is the Hartree potential, $V_{\mathrm{xc}}[\rho]$ is the relativistic exchange and correlation potential, and $V_{\mathrm{ext}}$ is the external potential due to the interaction of the electrons with the nucleus.

The nuclear charge distribution $\rho_{\mathrm{nuc}}(r)$ is modeled by using a two-parameter Fermi-type distribution \cite{Palffy10}. The value of the root mean square charge radius of the nucleus, which is needed as an input parameter, is taken from Ref. \cite{ANGELI2013} where existing experimental data are quoted. Otherwise, the empirical formula from Ref. \cite{JOHNSON1985405} is used. The external potential $V_{\mathrm{ext}}$ is then computed thanks to
	\begin{equation}
		V_{\mathrm{ext}}(r) = -\int \dfrac{\rho_{\mathrm{nuc}}(r')}{\left|\vec{r}-\vec{r}' \right|}d\vec{r}' \text{.}
	\end{equation}

For $V_{\mathrm{xc}}[\rho]$, we have used different exchange-correlation functionals: Vosko, Wilk and Nusair (VWN) \cite{Vosko80}, Gunnarsson and Lundqvist (GL) \cite{Gun76} and Perdew and Wang (PW) \cite{Perd92}. We have also implemented the gradient-dependent exchange-correlation functional of van Leeuwen and Baerends (LB) \cite{Leeuwen94}. This functional is nonlocal and depends not only on the electron density but also on the electron-density gradient.

We have elaborated a computer code for generating electron energies and wave functions. It is based on the numerical schemes of Ref. \cite{CERTIK20131777} and is complemented by the incorporation of a realistic description of the nucleus and by the implementation of the optimized effective potential method. We have used an exponential grid well adapted for radius values within the area of the nucleus, which is essential for the modeling of the electron-capture process. The possibility to create atomic vacancies, as needed for the modeling of electron capture, has also been implemented.

We have first tested and benchmarked our code by reproducing the atomic energies from Kotochigova \textit{et al.} \cite{Kot97} for atoms from Hydrogen to Uranium. They have been obtained within the framework of RLDA using the Vosko, Wilk and Nusair exchange-correlation functional and a point-charge model for describing the charge distribution of the nucleus. 

\begin{figure}[!ht]
\begin{center}
\includegraphics[scale=0.15]{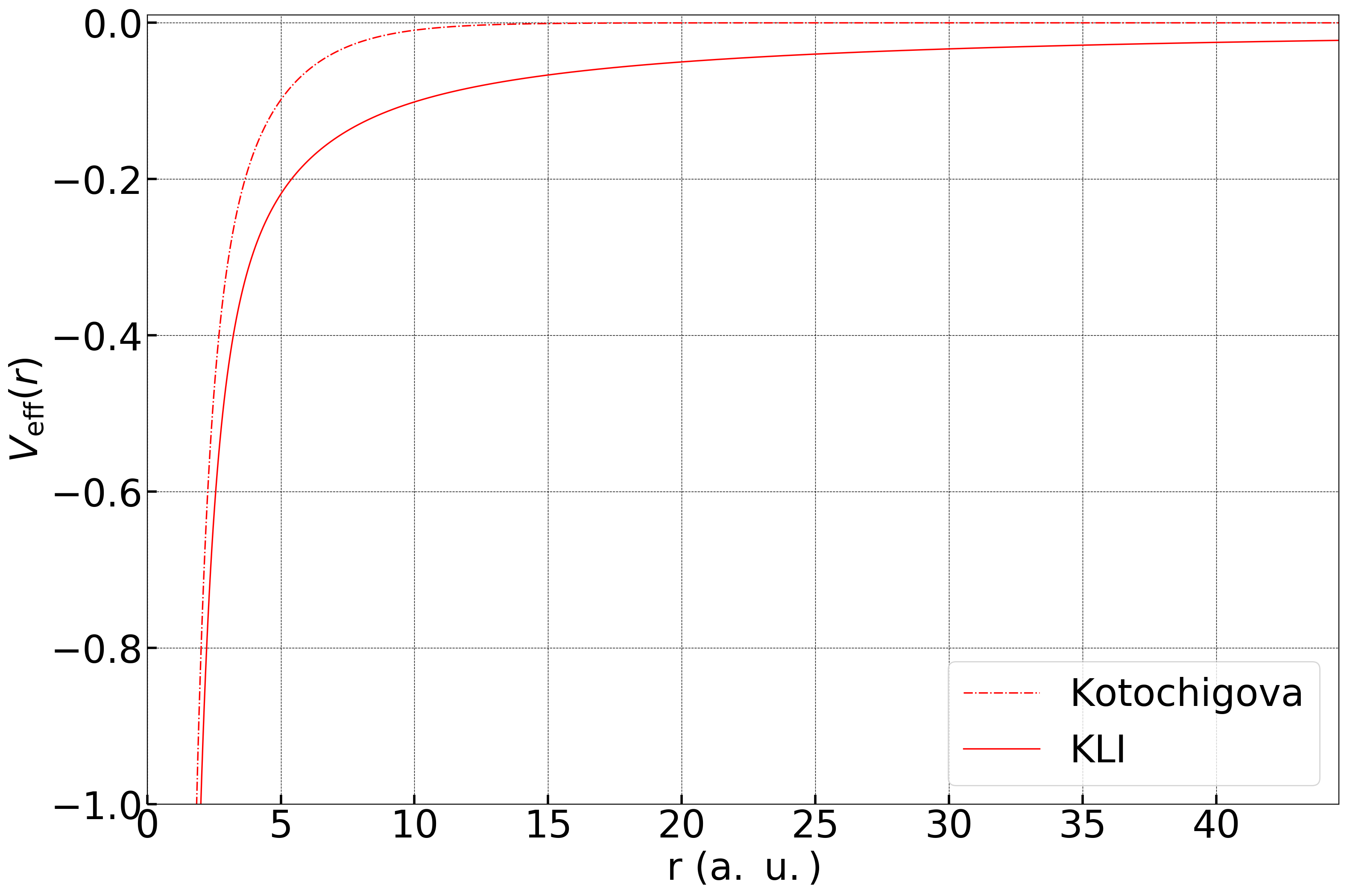}
\caption{KLI (solid) and Kotochigova (dashed) effective potentials for \rn{109}{Cd}.}
\label{fig:pot109Cd} 
\end{center}
\end{figure}

In a second step, we have optimized our modeling in regards to the effective potential. For a neutral atom, the asymptotic behavior of the effective potential is given by the exchange contribution $V_{\mathrm{xc}}$, which behaves at large distance as $\rho^{1/3}$. As a consequence, the effective potential $V_{\mathrm{eff}}$ decreases exponentially to zero and does not reproduce the expected $1/r$ asymptotic behavior. This problem does not appear in the Dirac-Fock theory because the Dirac-Fock exchange potential exactly compensates the self-interaction term contained in the Hartree potential (see Fig. \ref{fig:pot109Cd}).
	
Generally, the use of this correction leads to a much better description of the outermost shells. In particular, the experimental ionization potential can be well reproduced \cite{Leeuwen94}. In addition, this correction is crucial for a correct description of unoccupied and continuum states that are essential for shake-up and shake-off processes. However, the drawback of this correction and its variants (ADSIC \cite{CIOFINI200567}, SIC \cite{Perdew81}, GAM \cite{Politis98}) is an inaccurate description of the inner shells, which are mainly involved in the electron-capture process (i.e., $1s_{1/2}$, $2s_{1/2}$, $2p_{1/2}$). A way to overcome this inconvenience is to use the optimized effective potential method, which was originally described within a nonrelativistic framework by J. B. Krieger \textit{et al.} \cite{Kri92,Kri93} and extended to the relativistic case by Xiao-Min Tong and Shih-I Chu \cite{Tong98}.

To compute the optimized effective potential $V_{\mathrm{eff}}^{\mathrm{KLI}}$ -- giving what we call the KLI model in the following, one has to solved self-consistently the equation
	
	\begin{widetext}
		\begin{equation}
			\begin{array}{ll}
				V_{\mathrm{eff}}^{\mathrm{KLI}}(r) = &V_{\mathrm{eff}}[\rho](r) +\dfrac{N_{N_{\mathrm{orb}}}\rho_{N_{\mathrm{orb}}}(r)}{\rho(r)} \times \left(-V_{\mathrm{H}}[\rho_{N_{\mathrm{orb}}}](r) - V_{\mathrm{xc}}[\rho_{N_{\mathrm{orb}}}](r)\right) \\ 
				\\
				& + \mysum{i=1}{N_{\mathrm{orb}}-1}{\dfrac{N_i\rho_i(r)}{\rho(r)}} \times \left\{ -V_{\mathrm{H}}[\rho_i](r) - V_{\mathrm{xc}}[\rho_i](r)-\int_0^{\infty} \rho_i(r) (-V_{\mathrm{xc}}[\rho_i](r) \right. \\
				\\
				& \left. - V_{\mathrm{H}}[\rho_i](r))r^2 dr + \int_0^{\infty} \rho_i(r) V_{\mathrm{eff}}^{\mathrm{KLI}}(r) r^2 dr \right\} \text{,}
			\end{array}
		\end{equation}
	\end{widetext}
where $N_i$ is the occupation number of the orbital $i$. The term $V_{\mathrm{eff}}^{\mathrm{KLI}}(r)$ that appears on both sides of the above equation will quickly converge after a few iterations. All the details concerning the derivation of this equation are given in Ref. \cite{Tong98}.
	
\section{Results}
\label{sec:gen_res}

We have carried out many calculations employing different exchange-correlation functionals (VWN, GL, PW and LB) and self-interaction-corrected models (ADSIC, SIC, GAM and KLI). It turned out that KLI is the best compromise for obtaining: \textit{i)} electron binding energies of inner subshells close to the experimental values (see Table \ref{tab:bind_ener}); and \textit{ii)} a long-range Coulomb behavior of the effective potential (see Fig. \ref{fig:pot109Cd}). Moreover, we have found that within the KLI model, the results depend only very slightly on the exchange-correlation functional utilized. Therefore, we have employed the VWN functional for the comparison of the results obtained with both Kotochigova and MCDF models.   
	
\subsection{Binding energies}

Binding energies of the inner subshells of the atoms considered in the present work are given in Table \ref{tab:bind_ener} for Kotochigova \cite{Kot97,NIST09}, KLI and MCDF models. They are compared with experimental values taken in Refs. \cite{Sevier79,Bearden67}. One must point out that MCDF energies for \rn{37}{Ar}, \rn{41}{Ca} and \rn{109}{Cd} have been computed in single configuration, without the inclusion of electron correlations. For \rn{7}{Be}, the binding energies obtained in the MCDF framework and including correlations (see explanations in \ref{sec:MCDF}) are also given in Table \ref{tab:bind_ener}. It is clear that KLI energies are much more accurate, which validates the approach developed to correct for the asymptotic behavior of the effective potential and to obtain accurate binding energies for the inner subshells.\\

\vspace*{-0.3cm}
	\begin{longtable}{c | c c c c }
		\caption{Binding energies (in atomic units) of the innermost orbitals for Kotochigova \protect\cite{Kot97,NIST09} (RLDA approximation, neutral atoms), KLI and MCDF models. Experimental values are from Ref. \protect\cite{Sevier79} for \rn{7}{Be} to \rn{55}{Fe} atoms (in gaseous or vapor state) and from Ref. \protect\cite{Bearden67} for \rn{109}{Ce} to \rn{138}{La} atoms (in elemental or oxide state).}
		\label{tab:bind_ener}\\
			\hline \hline
			Element & Kotochigova & KLI & MCDF & Experimental \\\hline
			&  & $1s_{1/2}$ &  & \\\hline
			\rule[0.4cm]{0cm}{0cm} ${^7}$Be & -3.856 & -4.815 & -4.733& -4.384 \\ 
			&  &  & $~$[-4.752]$^*$ &  \\ 
			\rule[0.4cm]{0cm}{0cm} ${^{37}}$Ar & -114.08 & -118.14 & -119.05 & -117.704 (11) \\
			\rule[0.4cm]{0cm}{0cm} ${^{41}}$Ca &  -144.40 &  -148.93 & -150.05 & -148.57 (7) \\
			\rule[0.4cm]{0cm}{0cm} ${^{54}}$Mn & -235.08 & -240.81 &  ---&-240.70 (6) \\
			\rule[0.4cm]{0cm}{0cm} ${^{55}}$Fe & -255.90 & -261.88 & ---& -261.802 (37)  \\ 
			\rule[0.4cm]{0cm}{0cm} ${^{109}}$Cd & -968.69 & -981.42 & -985.50& -981.619 (11) \\
			\rule[0.4cm]{0cm}{0cm} ${^{125}}$I & -1204.13 &  -1218.71 & ---& -1218.953 (15) \\
			\rule[0.4cm]{0cm}{0cm} ${^{138}}$La & -1414.24 &  -1430.37 & ---& -1430.453 (15) \\\hline
			&  & $2s_{1/2}$ &  & \\\hline
			\rule[0.4cm]{0cm}{0cm} ${^7}$Be & -0.206 & -0.377 & -0.309 & -0.343  \\ 
			&  &  & $~$[-0.349]$^*$ & \\ 
			\rule[0.4cm]{0cm}{0cm} ${^{37}}$Ar & -10.86 & -11.74 & -12.41& -11.9803 (33) \\
			\rule[0.4cm]{0cm}{0cm} ${^{41}}$Ca & -15.16 & -16.14 & -16.96 & -16.26 (7)  \\
			\rule[0.4cm]{0cm}{0cm} ${^{54}}$Mn & -27.22 & -28.38 & ---& -28.72 (6) \\
			\rule[0.4cm]{0cm}{0cm} ${^{55}}$Fe & -29.99 & -31.20 & ---& -31.49 (7)  \\ 
			\rule[0.4cm]{0cm}{0cm} ${^{109}}$Cd & -143.97 & -146.64 & -149.64& -147.659 (11)  \\
			\rule[0.4cm]{0cm}{0cm} ${^{125}}$I & -186.24 & -189.33 & ---& -190.659 (11)  \\
			\rule[0.4cm]{0cm}{0cm} ${^{138}}$La & -225.36 & -228.79  & ---& -230.282 (18) \\\hline
			&  & $2p_{1/2}$ &  & \\\hline
			\rule[0.4cm]{0cm}{0cm} ${^{37}}$Ar & -8.50  & -9.40  & -9.62 & -9.2075 (26)     \\
			\rule[0.4cm]{0cm}{0cm} ${^{41}}$Ca &  -12.375 & -13.394  & -13.721 & -13.252 (29)    \\
			\rule[0.4cm]{0cm}{0cm} ${^{54}}$Mn &   -23.368 & -24.606  & ---  & -24.357 (37) \\
			\rule[0.4cm]{0cm}{0cm} ${^{55}}$Fe &   -25.92 & -27.21  & --- & -26.937 (37)   \\ 
			\rule[0.4cm]{0cm}{0cm} ${^{109}}$Cd &  -134.23   & -137.12   &  -138.71 &  -136.965 (11)  \\
			\rule[0.4cm]{0cm}{0cm} ${^{125}}$I & -174.97  & -178.32  & ---  & -178.311 (11)  \\
			\rule[0.4cm]{0cm}{0cm} ${^{138}}$La & -212.81  & -216.51 & --- & -216.476 (15)    \\
			\hline \hline
			\multicolumn{5}{l}{\rule[0.4cm]{0cm}{0cm} $^*$With correlations.}
	\end{longtable}

\vspace*{-0.5cm}
\subsection{Shaking probabilities}
\label{sec:shake}

Beyond binding energies, we investigated shaking probabilities in order to validate our atomic modeling. The initial state is the relaxed parent atom. Two final states have been studied, from different processes that create a vacancy: 
\begin{enumerate}[i)]
  \item \textit{Photoionization or internal conversion} -- The final state corresponds to a parent ion, with a vacancy in the orbital where the electron is ejected. We denote this as the Frozen Orbital (FO) approximation.
  \item \textit{Electron capture} -- The final state corresponds to a daughter atom with the electronic configuration of the parent atom, but without the captured electron. We denote this as the Daughter Excited (DE) approximation.
\end{enumerate}

Our predictions have been compared with partial results in the literature for a wide range of elements: \rn{7}{Be}, \rn{37}{Ar}, \rn{41}{Ca}, \rn{54}{Mn}, \rn{55}{Fe}, \rn{109}{Cd}, \rn{125}{I} and \rn{138}{La}. These results were generated from nonrelativistic \textit{ab initio} modeling techniques for various shell vacancies and assuming sudden approximation, as reported in Refs. \cite{Crase79,PhysRevA.36.693,AGKochur_2002,AGKochur_2005}. Shaking probabilities for each subshell of the different elements are given in the Appendix, determined with BetaShape, KLI and MCDF models. It is noteworthy that the perturbative approach to account for the hole effect in BetaShape leads to identical shaking probabilities with FO and DE approximations. 

For photoionization, Mukoyama and Taniguchi \cite{PhysRevA.36.693} results are for 1$s$, 2$s$, and 2$p$ vacancies up to Kr. Kochur and Popov \cite{AGKochur_2005} results depend only on the principal quantum number, i.e., no distinction was made between the shaking probabilities after e.g., a 2$s$ or a 2$p$ vacancy. In addition, we had to extract their numbers from their graphs and the uncertainties we quote are simple estimates due to our extraction procedure. For the electron capture process, results from Crasemann \textit{et al.} \cite{Crase79} are only provided for vacancies in the $s$ shells and 16 elements fom N to Xe. Marked probabilities in the Appendix have been interpolated.

\begin{figure}[!htp]
	\includegraphics[scale=0.16]{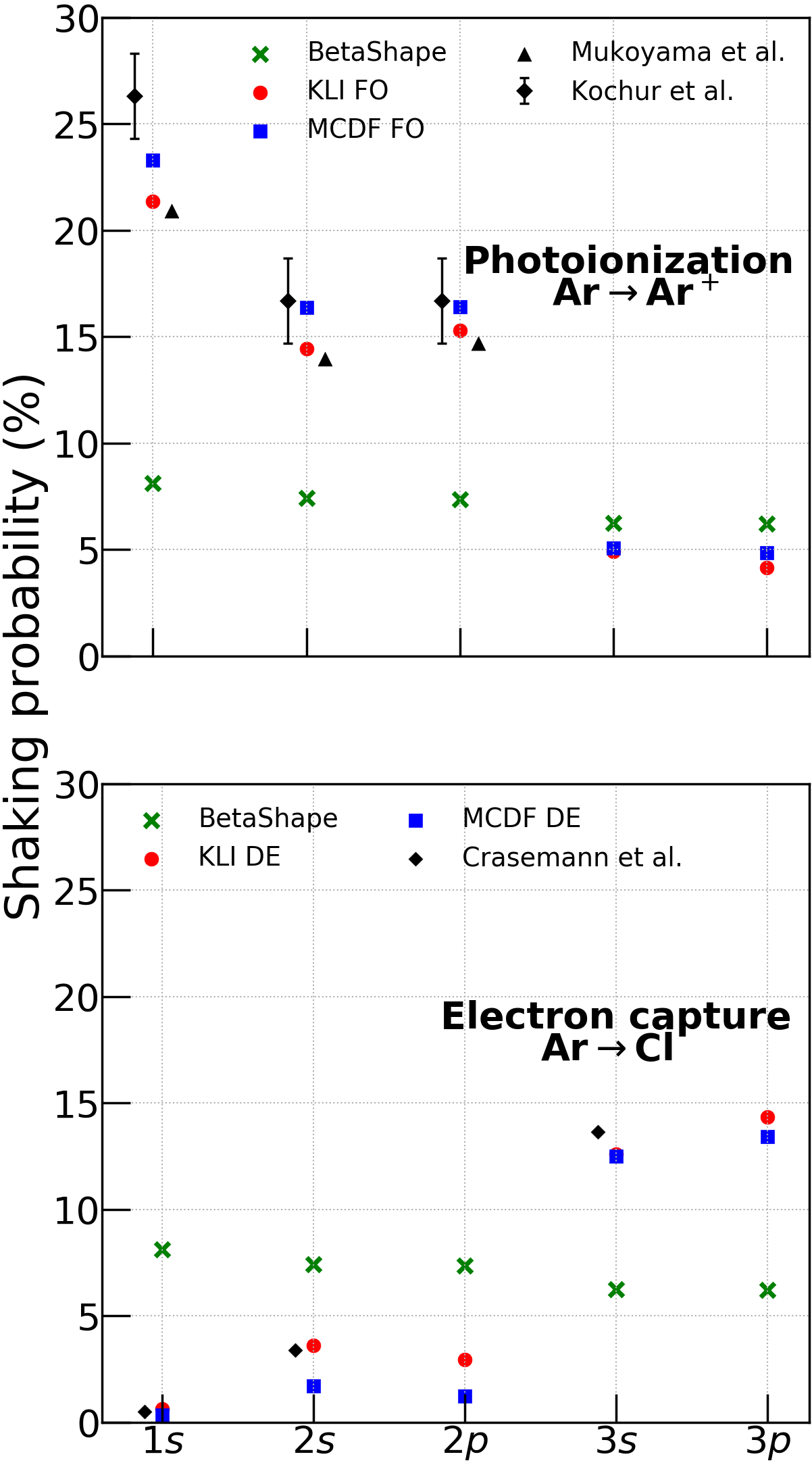}
	\vspace*{0.3cm}
	\caption{Shaking probability after a shell vacancy in \rn{37}{Ar} determined from three atomic models considered in this work. (\textit{Top}) Photoionization process, Frozen Orbital (FO) approximation. (\textit{Bottom}) Electron-capture process, Daughter Excited (DE) approximation. Nonrelativistic values are from Mukoyama \textit{et al.} \protect\cite{PhysRevA.36.693}, Kochur \textit{et al.} \protect\cite{AGKochur_2002} and Crasemann \textit{et al.} \protect\cite{Crase79}.}
	\label{fig:Shakes1}
\end{figure}

\begin{figure}[!htp]
	\includegraphics[scale=0.16]{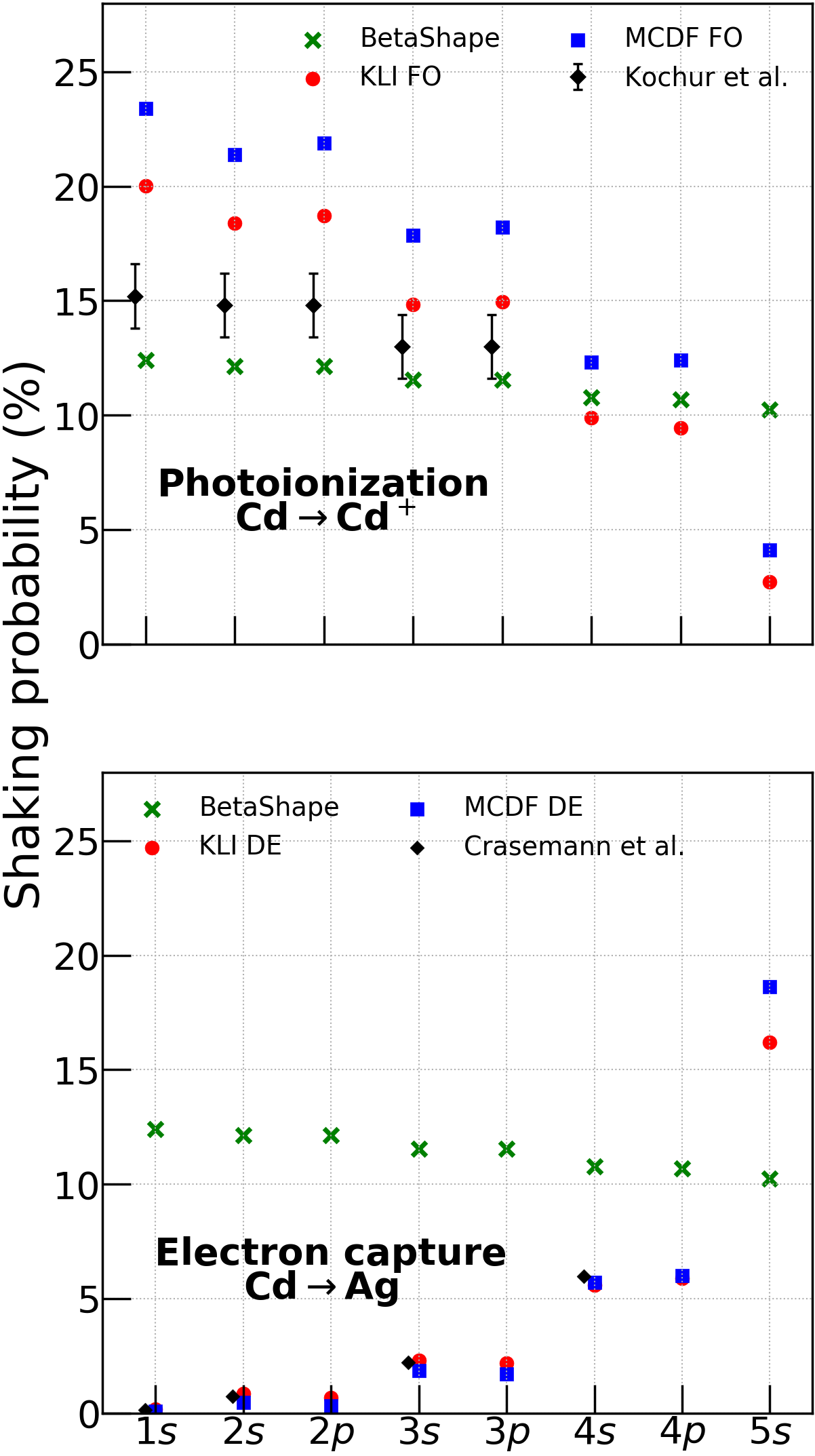}
	\vspace*{0.3cm}
	\caption{Shaking probability after a shell vacancy in \rn{109}{Cd} determined from three atomic models considered in this work. (\textit{Top}) Photoionization process, Frozen Orbital (FO) approximation. (\textit{Bottom}) Electron-capture process, Daughter Excited (DE) approximation. Nonrelativistic values are from Kochur \textit{et al.} \protect\cite{AGKochur_2002} and Crasemann \textit{et al.} \protect\cite{Crase79}.}
	\label{fig:Shakes2}
\end{figure}

We focus in this section on \rn{37}{Ar} and \rn{109}{Cd} predictions for their relevance and completeness regarding MCDF results. The shaking probabilities for different initial subshell vacancies are compared with the above-mentioned results in Figs.~\ref{fig:Shakes1} and~\ref{fig:Shakes2} for \rn{37}{Ar} and \rn{109}{Cd}, respectively.

\vspace*{0.5cm}
\subsubsection*{\rn{37}{Ar}} 
BetaShape model gives probabilities of 6\%-8\%, slightly dependent on the created vacancy. They are quite close to the KLI and MCDF predictions for the outer shells when considering the FO approximation. They strongly disagree in all other cases, either with KLI and MCDF predictions or with results from Refs. \cite{Crase79,PhysRevA.36.693,AGKochur_2005} for the FO and DE approximations, respectively. 

Remarkably, KLI and MCDF predictions for both processes are in good agreement with each other, but also with the results from Refs. \cite{Crase79,PhysRevA.36.693,AGKochur_2005}. In addition, both the KLI and MCDF methods exhibit a small dependency of the shaking probability on the vacancy subshell, which cannot be seen on the results from Refs. \cite{Crase79,PhysRevA.36.693,AGKochur_2005} because of their nonrelativistic approach.

\clearpage

\subsubsection*{\rn{109}{Cd}} 
As in \rn{37}{Ar} case, the shaking probabilities from BetaShape are quite constant (10\%-12\%) regardless of the vacancy location, and are close to KLI and MCDF predictions only for the outer shells (4$s$, 4$p$) with FO approximation. Considering electron capture, KLI and MCDF shaking predictions exhibit a striking agreement, and also with the probabilities from Ref. \cite{Crase79}. 

However for the photoionization process, MCDF results are systematically higher by about 3\% than KLI results. The values from Ref. \cite{AGKochur_2005} differ even more, which can be due in part to the imprecision of our graphical extraction method.

\subsubsection*{Discussion}
Our relativistic approach confirms that shaking is more likely for inner-shell vacancies when considering the FO approximation. 
In contrast, the DE approximation gives a higher probability of shaking when an outer shell vacancy is created.
This difference is due to the overlap between the initial and final atomic states involved, as can be seen from Eq. (\ref{eq:sudden_shakes}).
The closer the final state is to the initial state, the weaker the perturbation of the electron cloud, which results in a lower probability of shaking.

The physical process at the origin of the vacancy has thus a major impact on the shaking probabilities. They are related to the capture probabilities via Eq. (\ref{eq:lambda_ec}). Consequently, the atomic model employed to describe the initial and final states is critical to any realistic theoretical prediction. It is clear that the BetaShape model fails to provide accurate shaking probabilities, especially for the innermost subshells.

The quoted shaking probabilities from Kochur and Popov \cite{AGKochur_2005} are the sum of the shake-up and shake-off probabilities for electrons in the $L$, $M$ and $N$ shells after photoionization of an electron belonging to the $K$, $L$ or $M$ shell. The error due to this restriction on the origin of excited and ejected electrons is acceptable for light elements since only $K$ and $O$ shell are missing. Indeed, Figs.~\ref{fig:Shakes1} and~\ref{fig:Shakes2} and shaking probabilities in the Appendix show a difference of 10\%-50\% between our calculations and those from \cite{AGKochur_2005} up to Cd. However for heavier atoms such as lanthanum and iodine, the shaking probabilities for electrons from other shells than $L$, $M$ and $N$ contribute significantly to the overall sum, and the difference with our predictions is a factor of three to four.

\subsection{Capture probabilities}

The weak interaction transforms the parent nucleus with atomic number $Z$ into the daughter nucleus with atomic number ($Z$-1). This interaction is mediated by massive bosons and is therefore of very short range. The timescale of the electron-capture decay might be seen as instantaneous compared with the atomic timescale, i.e., the vacancy lifetime. We thus considered the DE approximation for the calculation of the capture probabilities.

We selected some radionuclides of interest to test the predictions of the presented atomic models under different conditions: atomic number ($Z=4-57$); transition nature (allowed: \rn{7}{Be}, \rn{37}{Ar}, \rn{54}{Mn}, \rn{55}{Fe}, \rn{109}{Cd} and \rn{125}{I}; first forbidden unique: \rn{41}{Ca}; second forbidden unique: \rn{138}{La}); and availability of accurate measurements to compare with (except for \rn{41}{Ca}). The dominant electron-capture transition in each case was studied. Calculation of capture probabilities were performed using the recommended $Q$ values established in the latest Atomic Mass Evaluation AME2020 \cite{Wang21}. Nuclear level energies were taken from the latest ENSDF evaluations for the following decays: \rn{54}{Mn} \cite{DONG20141}, \rn{109}{Cd} \cite{KUMAR20161}, \rn{125}{I} \cite{KATAKURA2011495} and \rn{138}{La} \cite{CHEN20171}. 

\begin{figure}[!htp]
	\includegraphics[scale=0.14]{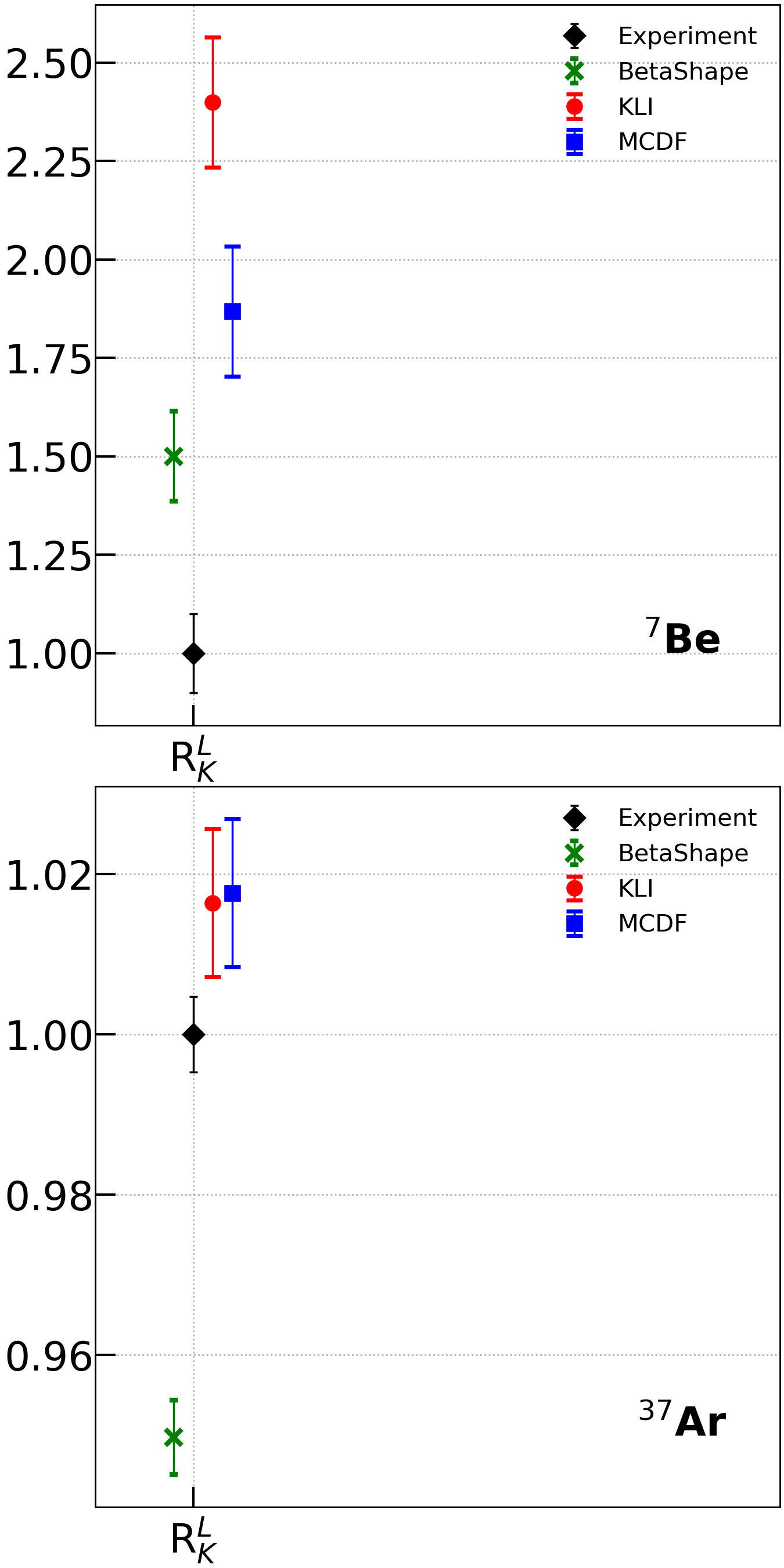}
	\vspace*{0.3cm}
	\caption{Capture probability ratios $R_K^L$ for \rn{7}{Be} (\textit{top}) and \rn{37}{Ar} (\textit{bottom}). Predictions of different atomic models are compared: KLI (circles), MCDF (squares) and BetaShape (crosses). Black diamonds indicate experimental uncertainties.}
    \label{fig:Capture1}
\end{figure}

\begin{figure}[!htp]
	\includegraphics[scale=0.14]{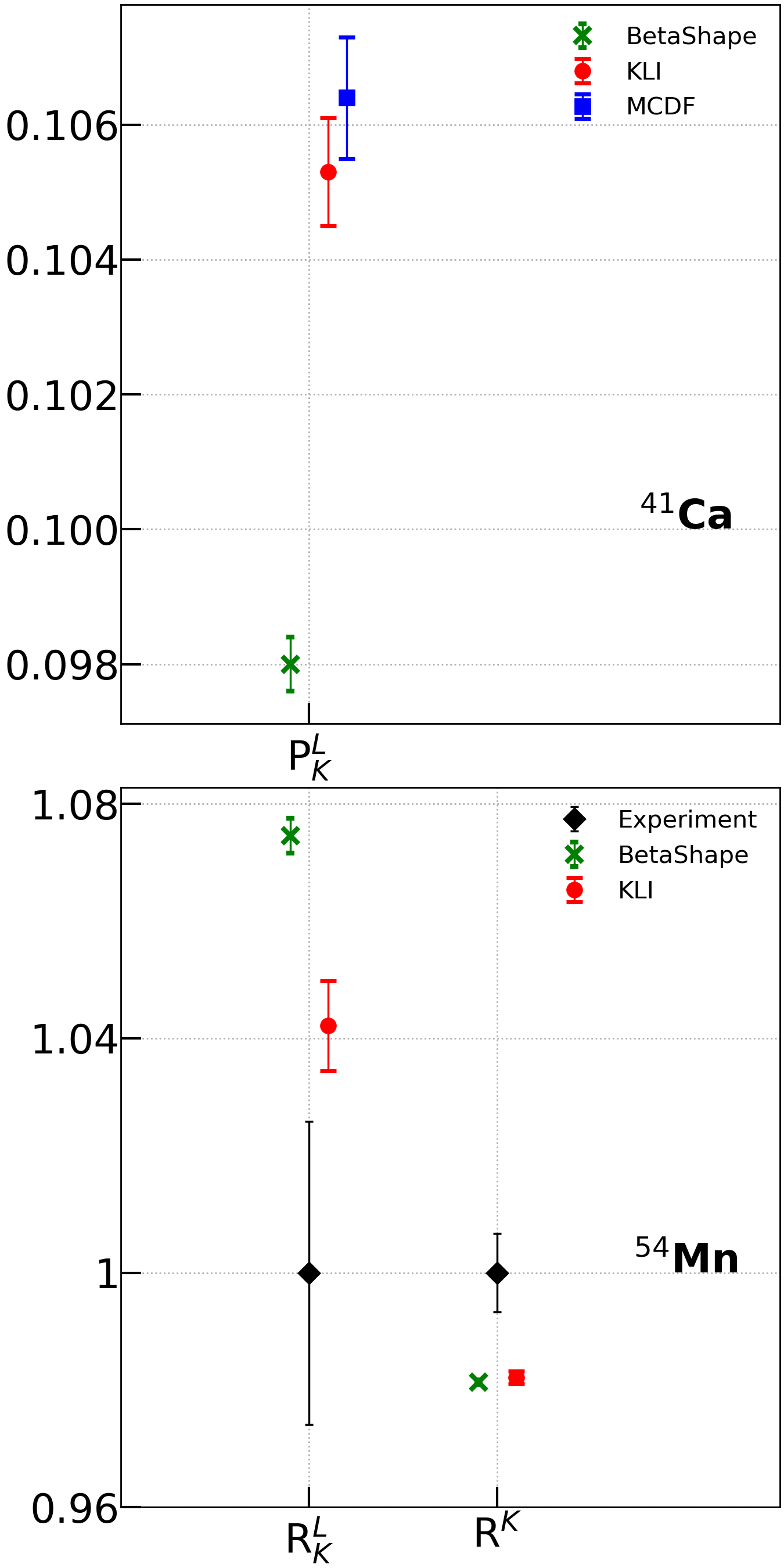}
	\vspace*{0.3cm}
	\caption{Capture probability ratios $P_L/P_K$ for \rn{41}{Ca} (\textit{top}) and $R_K^L$, $R^K$ for \rn{54}{Mn} (\textit{bottom}). The meaning of the symbols is the same as in Fig. \protect\ref{fig:Capture1}.}
	\label{fig:Capture2}
\end{figure}

\begin{figure}[!htp]
	\includegraphics[scale=0.14]{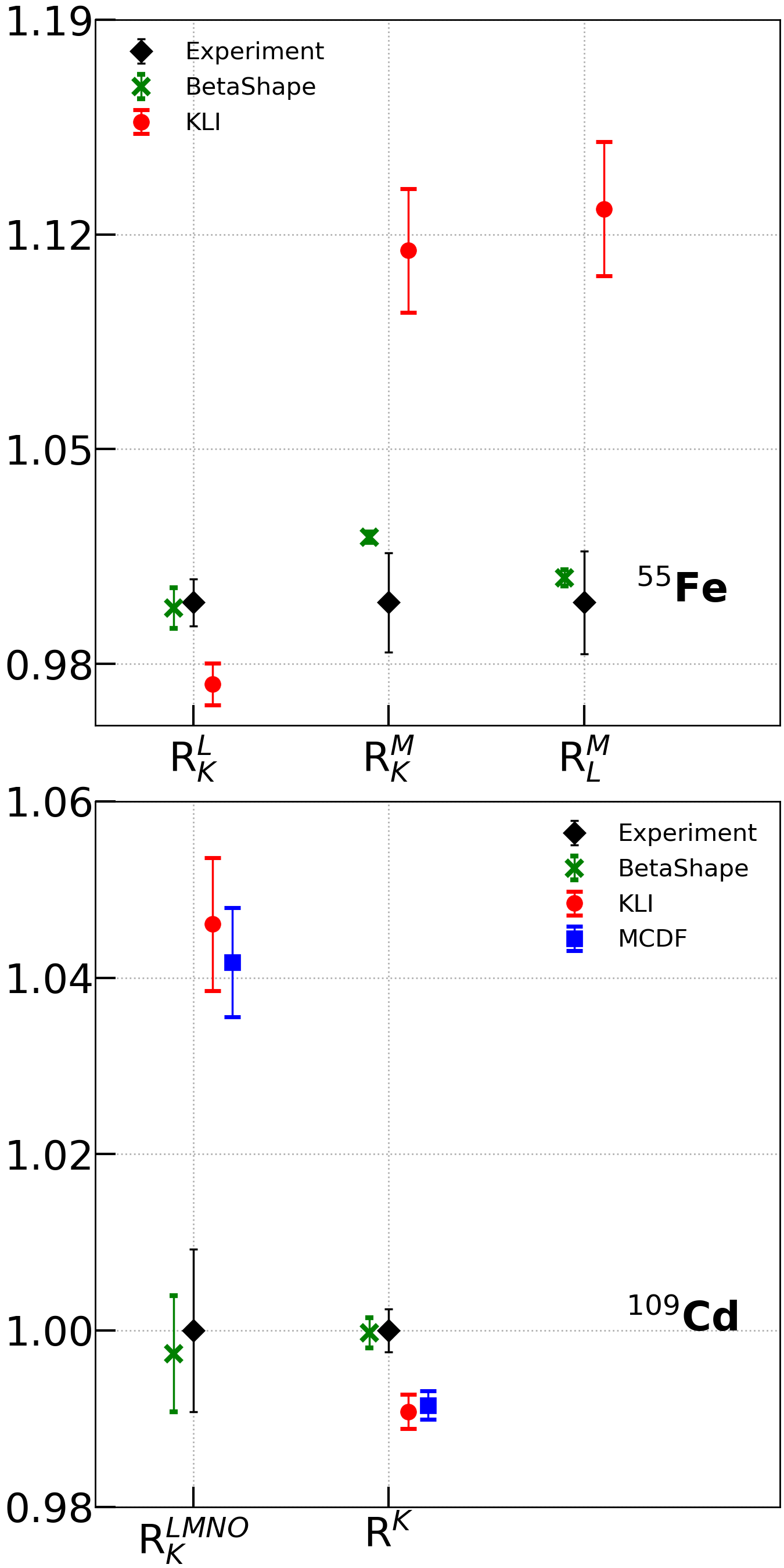}
	\vspace*{0.3cm}
	\caption{Capture probability ratios $R_K^L$, $R_K^M$ and $R_L^M$ for \rn{55}{Fe} (\textit{top}), $R_K^{LMNO}$ and $R^K$ for \rn{109}{Cd} (\textit{bottom}). The meaning of the symbols is the same as in Fig. \protect\ref{fig:Capture1}.}
	\label{fig:Capture3} 
\end{figure}

\begin{figure}[!htp]
	\includegraphics[scale=0.14]{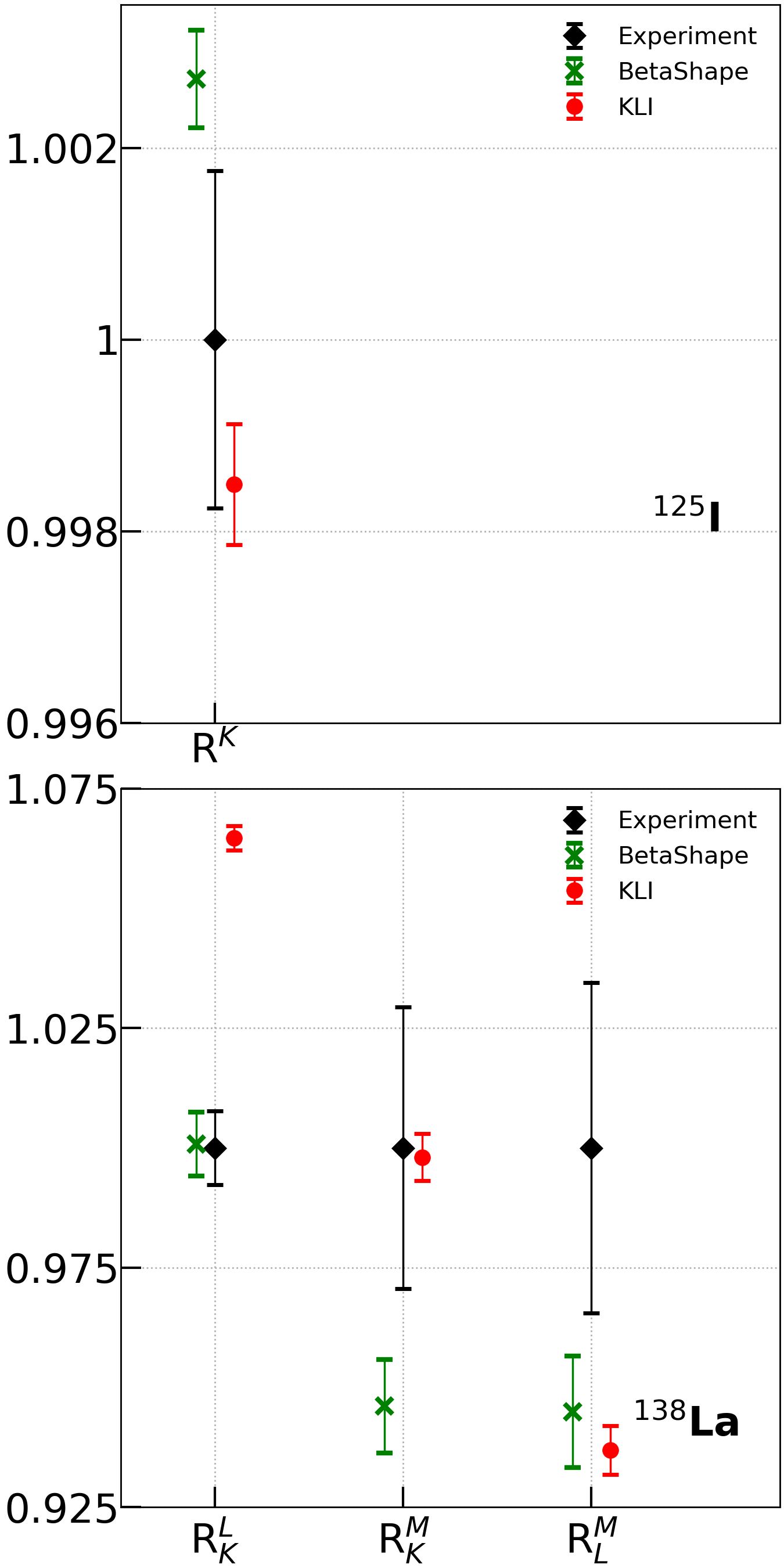}
	\vspace*{0.3cm}
	\caption{Capture probability ratios $R^K$ for \rn{125}{I} (\textit{top}), $R_K^L$, $R_K^M$, and $R_L^{M}$ for \rn{138}{La} (\textit{bottom}). The meaning of the symbols is the same as in Fig. \protect\ref{fig:Capture1}.}
	\label{fig:Capture4} 
\end{figure}

Most often, experimental values are given as relative, i.e., as a ratio of capture probabilities between two shells, instead of absolute capture probabilities, which are much more difficult to measure precisely. To unify the presentation of their comparison with the theoretical predictions, we defined the following quantities 
\begin{equation}
\label{eq:ec_ratios} 
\begin{array}{lcl}
    \mathrm{R}^i & = & \dfrac{P_i}{(P_i)_{\mathrm{exp}}} \text{,} \\
    && \\
    \mathrm{R}^i_j & = & \left(\dfrac{P_i}{P_j}\right) / \left(\dfrac{P_i}{P_j}\right)_{\mathrm{exp}} \text{,}
\end{array}
\end{equation}
where $P_i = P_{n \kappa}$ is defined in Eq. (\ref{eq:lambda_ec}). Numerator values are either from BetaShape, KLI, or MCDF models. Figures \ref{fig:Capture1} to \ref{fig:Capture4} show, for the selected radionuclides, the capture probability ratios obtained with the different models. Detailed data are also given in Table \ref{tab:capt_prob} and are discussed below.

\clearpage
\begin{center}
\begin{table*}
\caption{Comparison of calculated and measured capture probabilities for different isotopes considered in the present work. The three models and the experimental values are described in the text. \vspace*{0.3cm}}
\begin{ruledtabular}
\begin{tabular}{lllllll}
\rule[0.4cm]{0cm}{0cm}{Isotope} & Energy (keV) & Quantity & Experimental    & BetaShape & KLI & MCDF \\ \hline
\rule[0.4cm]{0cm}{0cm}\rn{7}{Be}   & 861.89 (7)  & $P_L/P_K$ & 0.070 (7)    & 0.105 (8)    & 0.168 (12)   &  $~$0.131 (10)  \\ 
   &  &  &  &  &  & [0.127 (11)]\footnote{With correlations.}  \\ \hline
\rule[0.4cm]{0cm}{0cm}\rn{37}{Ar}  & 813.87 (20) & $P_L/P_K$ & 0.09750 (46) & 0.09260 (45)   & 0.0991 (9)   &  0.09922 (26) \\ \hline
\rule[0.4cm]{0cm}{0cm}\rn{41}{Ca}  & 421.64 (14) & $P_L/P_K$ & --	    & 0.09800 (40) & 0.1053 (8)   &  0.1064 (9)   \\ \hline
\rule[0.4cm]{0cm}{0cm}\rn{54}{Mn}  & 542.3 (10)  & $P_L/P_K$ & 0.1044 (27)  & 0.11219 (31) & 0.1088 (8)   &  -- 	  \\ 
\rule[0.4cm]{0cm}{0cm}             &             & $P_K$     & 0.901 (6)    & 0.88419 (34) & 0.8849 (10)  &  -- 	  \\ \hline
\rule[0.4cm]{0cm}{0cm}\rn{55}{Fe}  & 231.12 (18) & $P_L/P_K$ & 0.1165 (9)   & 0.11629 (31) & 0.1134 (8)   &  -- 	  \\ 
\rule[0.4cm]{0cm}{0cm}             &             & $P_M/P_K$ & 0.01786 (29) & 0.01824 (12) & 0.01991 (36) &  -- 	  \\ 
\rule[0.4cm]{0cm}{0cm}             &             & $P_M/P_L$ & 0.1556 (26)  & 0.1568 (11)  & 0.1756 (34)  &  -- 	  \\ \hline
\rule[0.4cm]{0cm}{0cm}\rn{109}{Cd} & 127.1 (18) & $P_{LMNO}/P_K$ & 0.2279 (21)  & 0.2273 (15)  & 0.2384 (17)  &  0.2374 (14)  \\ 
\rule[0.4cm]{0cm}{0cm}             &   & $P_K$     & 0.815 (2)    & 0.8148 (14)  & 0.8075 (16)  &  0.8081 (13)  \\ \hline
\rule[0.4cm]{0cm}{0cm}\rn{125}{I}  & 150.28 (6)  & $P_K$     & 0.7971 (14)  & 0.79927 (41) & 0.7959 (5)   &  -- 	  \\ \hline
\rule[0.4cm]{0cm}{0cm}\rn{138}{La} & 312.59 (34) & $P_L/P_K$ & 0.391 (3)    & 0.3913 (26)  & 0.4163 (10)  &  -- 	  \\ 
\rule[0.4cm]{0cm}{0cm}             &             & $P_M/P_K$ & 0.102 (3)    & 0.0965 (10)   & 0.1018 (5)   &  -- 	  \\ 
\rule[0.4cm]{0cm}{0cm}             &             & $P_M/P_L$ & 0.261 (9)    & 0.2466 (30)  & 0.2445 (13)  &  -- 	  \\ 
\end{tabular}
\end{ruledtabular}
\label{tab:capt_prob}
\end{table*}
\end{center}

\vspace*{-1cm}
\subsubsection*{\rn{7}{Be}}
The experimental value of $P_L/P_K=0.070(7)$ is from a precise measurement in which \rn{7}{Be} was implanted in a thin Ta metallic film \cite{BeEST2020}. It is clear that none of the calculated values agrees and that the atomic modeling has a very strong influence on the $P_L/P_K$ ratio, with a factor of 1.5 to 2.4  between the predictions. The chemical form of \rn{7}{Be} was demonstrated to have a significant influence on its decay half-life, which seems reasonable with only two filled atomic orbitals. It is noteworthy that a first experiment made by Voytas \textit{et al.} \cite{PhysRevLett.88.012501} in a HgTe layer gave a significantly different ratio of $P_L/P_K=0.040(6)$. The $P_L/P_K$ ratio is thus expected to be influenced by environmental effects and the discrepancy with the theoretical predictions could be due the assumption of a decay in vacuum. In medium effects were estimated by Ray \textit{et al.} in Ref. \cite{TaMedium} with a simple model and a strong correction factor of 0.2986 was determined for metallic Ta, and of 0.577 for HgTe. Applying these corrections to our predictions reduces the disagreement but the former seems too high and the latter insufficient.

\subsubsection*{\rn{37}{Ar}}
The reference value of $P_L/P_K=0.09750(46)$ is a weighted average of the measurements given in the review from Bambynek \textit{et al.} \cite{Bamb77}. KLI and MCDF results perfectly agree and both are much closer to the reference value (consistent at 2$\sigma$) than BetaShape result, which underestimates the $P_L/P_K$ ratio by 5\%.

\subsubsection*{\rn{41}{Ca}}
This transition is first forbidden unique, the only one considered in the present work. In this case, $C_{n\kappa}$ in Eq. (\ref{eq:C_nk}) is not equal to unity and must be taken into account. To our knowledge, no experimental value of any capture probability is available in the literature for this radionuclide. We can simply observe that the $P_L/P_K$ ratios from KLI and MCDF agree well and are consistent within the uncertainties. The BetaShape value is lower by about 7\%, as for \rn{37}{Ar} decay.

\subsubsection*{\rn{54}{Mn}}
The first experimental value comes from the weighted mean of two old measurements with quite large uncertainties: $P_L/P_K =$ 0.098 (6) from Ref. \cite{Moler63} and $P_L/P_K =$ 0.106 (3) from Ref. \cite{Manduchi63}. 

The experimental $P_K$ probability has been determined examining the measured values listed in Ref. \cite{Bamb77,DDEP_54Mn}. Only two consistent measurements do not depend on the fluorescence yield $\omega_K$(Cr) and we chose for our comparison their weighted mean $P_K =$ 0.901 (6). Nine other measurements do depend on $\omega_K$(Cr) because the experimental techniques employed did not allow to detect both the emitted x rays and Auger electrons. Their weighted mean is $P_K \omega_K =$ 0.253 (6). It is interesting to note that the fluorescence yield $\omega_K \text{(Cr)} =$ 0.289 (5) from a semi-empirical fit from Bambynek \cite{Bamb84} is significantly different from the experimental value $\omega_K \text{(Cr)} =$ 0.2793 (17) from the same author \cite{Bamb67om}. Using the former leads to $P_K =$ 0.875 (26) while with the latter, we have deduced a higher value $P_K =$ 0.906 (22). This suggests that the correct fluorescence yield should be close to the experimental value. 

BetaShape and KLI $P_K$ probabilities agree well with each other, and are about 2\% lower than the experimental value. However, the predictions are in good agreement with the $P_K$ value established with the fluorescence yield from the semi-empirical fit. Regarding the $P_L/P_K$ ratio, KLI result is much closer to the experimental value than BetaShape prediction, which is about 8\% higher. 

\subsubsection*{\rn{55}{Fe}}
All three experimental capture probability ratios result from the weighted means of two precise measurements \cite{Pengra72,Loidl18}. The BetaShape predictions are in excellent agreement, consistent within the uncertainties. Surprisingly, KLI results deviate much more: $P_L/P_K$ is lower by about 3\%; $P_M/P_K$ and $P_M/P_L$ are higher by about 10\%.

\subsubsection*{\rn{109}{Cd}}
The experimental values have been determined as weighted means of several measurements listed in Ref. \cite{DDEP_109Cd}. The $P_{LMNO}/P_K$ value has been reevaluated by revising one of the measured values as suggested in Ref. \cite{Bamb77}. The BetaShape model provides the most accurate predictions, fully consistent with experiment, with less than 0.3\% difference. KLI and MCDF results are in very good agreement with each other, but it seems that they fail to reproduce the experimental values. It is true for the $P_{LMNO}/P_K$ ratio, which is 4\% higher than expected for both. However, the $P_K$ probabilities are less than 1\% lower and only appear off because of the small relative experimental uncertainty (0.2\%).

\subsubsection*{\rn{125}{I}}
Seven measurements of the $P_K$ capture probability were performed in the past, listed in Ref. \cite{Bamb77,DDEP_125I}. The experimental value in Table \ref{tab:capt_prob} results from their weighted mean. The value is largely dominated by the measurement from Ref. \cite{Leutz64} and only a slight dependency in the $\omega_K$(Te) fluorescence yield is expected. KLI prediction is consistent with the experimental reference, an exceptional agreement considering the small relative uncertainty (0.2\%) of the latter. BetaShape result is not consistent but exhibits a disagreement of only 0.3\%.

\subsubsection*{\rn{138}{La}}

This transition is of second forbidden unique nature, the only one studied in the present work. As for \rn{41}{Ca} decay, $C_{n\kappa}$ in Eq. (\ref{eq:C_nk}) is not equal to unity and must be taken into account. All the measured values come from a single recent experiment \cite{Quarati16}. Comparing with theoretical predictions, the situation is quite unclear. For the $P_L/P_K$ ratio, the BetaShape result is in remarkable accordance with the experimental value, with less than 0.1\% difference. However, KLI results is higher by 6\%. It is the contrary for the $P_M/P_K$ ratio, for which BetaShape result is lower by 6\% while KLI result is in excellent agreement with experiment, with 0.2\% difference. For the $P_M/P_L$ ratio, BetaShape and KLI results agree with each other but any model provides an accurate prediction, both being lower by $\sim$6\%.

\subsubsection*{Discussion}
KLI and MCDF atomic models provide very consistent predictions of capture probabilities, except for \rn{7}{Be} decay. As discussed, such a low-mass nucleus was proved to be sensitive to its chemical environment. However, the difference might be also due to correlation effects. MCDF results in Table \ref{tab:capt_prob} have been determined without correlation between the electrons, while KLI model includes them through the effective potential. We have performed additional MCDF calculations with full correlation treatment. The results differ by only 3\% to those without correlations and remain consistent: $P_L/P_K =$ 0.127 (11) instead of 0.131 (10). The total shaking probabilities differ by about 7\%: 0.10789 instead of 0.11553 for a 1$s$ vacancy, and 0.46798 instead of 0.44076 for 2$s$ vacancy. Such differences are expected to be much smaller for higher $Z$. Therefore, electron correlations should not play a significant role in the context of this study.

\section{Conclusion and perspectives}

The two international collaborations that recommend nuclear decay data (DDEP and ENSDF) employ the BetaShape code in their evaluations in order to improve the accuracy of the beta and electron-capture properties. In the present work, we have studied in minute detail the influence of the modeling of the atomic electrons on the electron-capture process. 

A realistic model has been developed within the framework of RDFT, in which we have implemented a few exchange-correlation functionals and self-interaction-corrected models, among the most popular in the community. From our analysis, the best choice that fulfills some physical constraints -- binding energies of inner subshells, asymptotic behavior of the effective potential -- is provided by the optimized effective potential method originally developed by Krieger, Li and Iafrate (KLI). It turns out that within this model, correlation effects seem to play a minor role. This has been confirmed by comparing the predictions for \rn{7}{Be} decay with MCDF calculations with and without correlations. We found differences of only a few percent, and this effect should be most significant for low-mass nuclei. A more precise study with MCDF for medium-mass nuclei would require a very heavy computational burden.

The electron-capture model of BetaShape has been adapted in order to consistently use the binding energies, wave functions and electronic configurations from the KLI and MCDF approaches. It allowed us to precisely include the hole effect on the other orbitals due to the vacancy created by the capture process, while this effect is accounted for approximately with the BetaShape atomic modeling. 

The total shaking probabilities in both photoionization and electron-capture processes have been calculated using the FO and DE approximations, respectively. Due to a lack of measured values, the results of the three atomic models employed -- BetaShape, KLI and MCDF -- have been compared with available predictions \cite{Crase79,PhysRevA.36.693,AGKochur_2002,AGKochur_2005}, all established within nonrelativistic frameworks. The BetaShape atomic model cannot differentiate photoionization from electron capture in shaking calculations, and the predictions are not accurate. KLI and MCDF results reasonably agree for photoionization, also with Mukoyama and Taniguchi \cite{PhysRevA.36.693} and Kochur and Popov \cite{AGKochur_2005} results for low-mass nuclei. For electron capture, KLI and MCDF results are in very good agreement with those from Crasemann \textit{et al.} \cite{Crase79}. Our relativistic approach makes us confident in predicting realistic shaking probabilities for medium and high-$Z$ nuclei. Besides relativistic effects, our study also improves the description of the shaking process by considering all possible vacancy creation scenarios and all subshell dependencies. 

The theoretical capture probabilities of several transitions of interest have been compared with experimental values with relative uncertainties from 0.2\% to 3.5\%, except for \rn{7}{Be} (10\%) and \rn{41}{Ca} (no existing measurement). Such a comparison covers a wide range of atomic numbers, $3 \leq Z \leq 57$, as well as different transition natures. KLI and MCDF predictions agree well and are better than BetaShape results for \rn{37}{Ar}, \rn{54}{Mn}, \rn{125}{I} and the $P_M/P_K$ ratio in \rn{138}{La} decay. BetaShape predictions are surprisingly in much better agreement with experiment in all other cases. Our understanding is that the inaccuracies of its atomic model -- binding energies, hole and shaking effects -- somehow compensate each other. However, it is not always true as clearly seen with \rn{37}{Ar}, \rn{41}{Ca} and \rn{54}{Mn} decays, without any hint to anticipate such a breakdown. New high-precision measurements are needed to explore this in detail, with more complete set of capture probabilities per radionuclide that include outer shells.

In the near future, the KLI atomic model will be extended to the continuum states, which will allow the separate computation of shake-up and shake-off processes. Indeed, good description of the unoccupied levels and the continuum states is mandatory for such processes, shake-off being in particular conditioned by the asymptotic behavior of the effective potential. In addition, this will allow us to determine capture-to-positron ratios, for which numerous precise measurements exist \cite{Bamb77,Mougeot19}.

The electron-capture model should also benefit from ongoing developments that aim at including a realistic nuclear structure for the calculation of forbidden nonunique transitions. Such an improvement will necessitate a revision of the overlap and exchange correction, with a precise coupling of lepton and nucleon wave functions.
	
Finally, another important issue concerns the role played by the environment on the electron-capture process, as seen in \rn{7}{Be} decay. Such a complicated problem deserves in itself a detailed specific study. 	

\begin{acknowledgments}

The authors would like to acknowledge the High Performance Computing Center of the University of Strasbourg for supporting this work by providing scientific support and access to computing resources. This work has received funding from the EMPIR projects 17FUN02 MetroMMC and 20FUN04 PrimA-LTD, co-financed by the Participating States and from the European Union’s Horizon 2020 research and innovation program. This research was also funded in part by the Fundação para a Ciência e Tecnologia (FCT, Portugal) through research center Grant No. \href{https://doi.org/10.54499/UIDB/04559/2020}{UIDB/04559/2020} to LIBPhys-UNL. Finally this work was supported by the French National Research Agency (ANR) through the Programme d'Investissement d'Avenir under Contract No. ANR-17-EURE-0024.

\end{acknowledgments}

\clearpage
\appendix

\section*{Appendix: Shaking probabilities}
\label{sec:Shaking_data}

We provide in Table \protect\ref{tab:shake} of this Appendix the probabilities of electron displacement following the creation of a vacancy calculated for different subshells of various elements. The vacancy is created either by photoionization or electron-capture process. BETASHAPE atomic modeling cannot distinguish between the two processes. KLI and MCDF approaches employed the frozen orbital (FO) approximation for the former, and the daughter excited (DE) approximation for the latter. Comparison is made with available predictions generated from nonrelativistic \textit{ab initio} modeling techniques for various shell vacancies and assuming sudden approximation (see section \protect\ref{sec:shake}).

\vspace*{2cm}
\begin{longtable}{lll|llll|lll}
	\caption{Shaking probabilities consecutive to the creation of an atomic vacancy in a given subshell due to photoionization or electron capture. See text for detailed explanations.}\label{tab:shake}\\
			\hline \hline
			\multicolumn{3}{c}{} & \multicolumn{4}{|c|}{Photoionization} & \multicolumn{3}{c}{Electron capture} \\ \hline
			\multirow{2}{*}{Parent} & \multirow{2}{*}{Vacancy} & \multirow{2}{*}{BetaShape} & KLI  & MCDF & Mukoyama \& & Kochur \& & KLI & MCDF  & Crasemann \\ 
			& &  & (FO)  & (FO) & Taniguchi \protect\cite{PhysRevA.36.693} & Popov \protect\cite{AGKochur_2005} & (DE) & (DE)  & \textit{et al.} \protect\cite{Crase79}  \\ \hline     
			
			\multirow{2}{*}{$^{7}_{4}$Be} & 1$s$ & 0.36649 & 0.17952 & 0.21779 & 0.21228 & 0.281 (14) & 0.09680 & 0.11553 & 0.05820$^*$  \\ 
			& &  &  & $~$[0.22112]$^{\dagger}$ &  &  &  & $~$[0.10789]$^{\dagger}$  &  \\ 
			& 2$s$ & 0.35005 & 0.04584 & 0.06530 & 0.03450 & - & 0.36171 & 0.44076 & 0.44716$^*$ \\ 
			& &  & & $~$[0.04797]$^{\dagger}$ & & & & $~$[0.46798]$^{\dagger}$ & \\ \hline
			
			\multirow{5}{*}{$^{37}_{18}$Ar} & 1$s$ & 0.08117 & 0.21336 & 0.23284 & 0.20918 & 0.263 (20) & 0.00628 & 0.00320 & 0.00489 \\ 
			& 2$s$ & 0.07420 & 0.14426 & 0.16360 & 0.13951 & \hspace*{-0.3cm}\rdelim\}{2}{*}[~0.167 (20)] & 0.03615 & 0.01714 & 0.03389 \\ 
			& 2$p_{1/2}$ & 0.07353 & 0.15303 & 0.16394 & 0.14673 &  & 0.02946 & 0.01218 & - \\ 
			& 3$s$ & 0.06248 & 0.04917 & 0.05080 & - & - & 0.12595 & 0.12491 & 0.13646 \\ 
			& 3$p_{1/2}$ & 0.06211 & 0.04140 & 0.04825 & - & - & 0.14316 & 0.13420 & - \\ \hline
			
			\multicolumn{10}{l}{~}  \\ 
			\multicolumn{10}{l}{~}  \\ 
			\multicolumn{10}{l}{~}  \\ 
			\multicolumn{10}{l}{~}  \\ 


			\hline
			\multirow{8}{*}{$^{41}_{20}$Ca} & 1$s$ & 0.27494 & 0.26792 & 0.36678 & 0.30259 & 0.381 (17) & 0.00549 & 0.00419 & 0.00452 \\ 
			& 2$s$ & 0.26945 & 0.20650 & 0.28849 & 0.24009 & \hspace*{-0.3cm}\rdelim\}{3}{*}[~0.310 (17)]  & 0.03095 & 0.03057 & 0.02939 \\ 
			& 2$p_{1/2}$ & 0.26864 & 0.21406 & 0.29366 & \hspace*{-0.3cm}\rdelim\}{2}{*}[~0.24714] & & 0.02502 & 0.02494 & - \\ 
			& 2$p_{3/2}$ & 0.26839 & 0.21327 & 0.29503 & &  & 0.02536 & 0.02295 & - \\ 
			& 3s & 0.25943 & 0.12799 & 0.19207 & - & \hspace*{-0.3cm}\rdelim\}{3}{*}[~0.185 (10)]& 0.10196 & 0.12612 & 0.10691 \\
			& 3$p_{1/2}$ & 0.25635 & 0.12191 & 0.19400 & - &  & 0.10607 & 0.11983 & - \\ 
			& 3$p_{3/2}$ & 0.25635 & 0.12082 & 0.19126 & - &  & 0.10687 & 0.11948 & - \\ 
			& 4$s$ & 0.24671 & 0.04416 & 0.07589 & - & - & 0.25329 & 0.34868 & -  \\ 
				
				
			\hline \multirow{6}{*}{$^{54}_{25}$Mn} & 1s & 0.33169 & 0.23974 & - & 0.25869 & 0.290 (17) & 0.00462 & - & 0.00341$^*$ \\
			& 2$s$ & 0.32739 & 0.20970 & - & 0.22538 & \hspace*{-0.3cm}\rdelim\}{2}{*}[ 0.274 (17)] & 0.02624 & - & 0.02171$^*$ \\ 
			& 2$p_{1/2}$ & 0.32687 & 0.21424 & - & 0.23048 & & 0.02138 & - & - \\ 
			& 3$s$ & 0.32074 & 0.12658 & - & - & \hspace*{-0.3cm}\rdelim\}{2}{*}[~0.125 (10)] & 0.06650 & - & 0.07134$^*$ \\ 
			& 3$p_{1/2}$ & 0.31905 & 0.12266 & - & - &  & 0.06758 & - & - \\ 
			& 4$s$ & 0.31313 & 0.03505 & - & - & - & 0.21349 & - & 0.25110$^*$ \\ \hline
			
			\multirow{6}{*}{$^{55}_{26}$Fe} & 1s & 0.03802 & 0.23136 & - & 0.24975 & 0.307 (17) & 0.00435 & - & 0.00380 \\
			& 2$s$ & 0.03394 & 0.20378 & - & 0.21894 & \hspace*{-0.3cm}\rdelim\}{2}{*}[~0.280 (17)] & 0.02456 & - & 0.02382 \\ 
			& 2$p_{1/2}$ & 0.03360 & 0.20807 & - & 0.22382 &  & 0.02001 & - & - \\ 
			& 3$s$ & 0.02692 & 0.12378 & - & - & \hspace*{-0.3cm}\rdelim\}{2}{*}[~0.120 (10)] & 0.06204 & - & 0.06341 \\ 
			& 3$p_{1/2}$& 0.02594 & 0.12028 & - & - &  & 0.06286 & - & - \\
			& 4$s$ & 0.02283 & 0.03414 & - & - & - & 0.20232 & - & 0.22058 \\ \hline
			
			\multirow{8}{*}{$^{109}_{48}$Cd} & 1s & 0.12395 & 0.20005 & 0.23381 & - & 0.152 (14) & 0.00161 & 0.00075 & 0.00131 \\
			& 2$s$ & 0.12152 & 0.18386 & 0.21374 & - &  \hspace*{-0.3cm}\rdelim\}{2}{*}[~0.148 (14)] & 0.00844 & 0.00456 & 0.00739 \\
			& 2$p_{1/2}$ & 0.12137 & 0.18717 & 0.21886 & - &  & 0.00676 & 0.00299 & - \\
			& 3$s$ & 0.11554 & 0.14831 & 0.17856 & - &  \hspace*{-0.3cm}\rdelim\}{2}{*}[~0.130 (14)] & 0.02311 & 0.01845 & 0.02201 \\
			& 3$p_{1/2}$ & 0.11531 & 0.14954 & 0.18203 & - &  & 0.02192 & 0.01709 & - \\
			& 4$s$ & 0.10770 & 0.09894 & 0.12317 & - & - & 0.05589 & 0.05717 & 0.05980 \\
			& 4$p_{1/2}$ & 0.10685 & 0.09436 & 0.12394 & - & - & 0.05878 & 0.06015 & - \\
			& 5$s$ & 0.10230 & 0.02706 & 0.04101 & - & - & 0.16187 & 0.18639 & - \\ \hline
			
			\multicolumn{10}{l}{~}  \\ 
            \multicolumn{10}{l}{~}  \\ 
            \multicolumn{10}{l}{~}  \\ 

			\hline \multirow{9}{*}{$^{125}_{53}$I} & 1s & 0.08764 & 0.21104 & - & - & 0.063 (10) & 0.00132 & - & 0.00107$^*$ \\ 
			& 2$s$ & 0.08518 & 0.18969 & - & - &  \hspace*{-0.3cm}\rdelim\}{2}{*}[~0.053 (10)] & 0.00678 & - & 0.00628$^*$ \\ 
			& 2$p_{1/2}$ & 0.08508 & 0.19405 & - & - &  & 0.00536 & - & - \\
			& 3$s$ & 0.07867 & 0.15781 & - & - & \hspace*{-0.3cm}\rdelim\}{2}{*}[~0.039 (10)] & 0.01961 & - & 0.01840$^*$ \\ 
			& 3$p_{1/2}$ & 0.07859 & 0.15932 & - & - & & 0.01846 & - & -  \\ 
			& 4$s$ & 0.06853 & 0.11863 &- &- &- & 0.04710 &- & 0.05015$^*$ \\ 
			& 4$p_{1/2}$ & 0.06831 & 0.11643 &-	&- &- & 0.04865 &-	& -  \\ 
			& 5$s$	& 0.05841 & 0.05211 &- & - &- & 0.11458 &-	&-  \\
			& 5$p_{1/2}$ & 0.05888 & 0.03975 &-	&-	&-	&0.13501 & - &-  \\ 
				

			\hline \multirow{19}{*}{$^{138}_{57}$La} &	1$s$& 0.08202 &	0.26099 &	-&	-&	0.041 (10) &	0.00124& -&	0.00095$^*$ \\
			& 2$s$ & 0.07944 & 0.24204 &- & - & \hspace*{-0.3cm}\rdelim\}{3}{*}[~0.039 (10)] & 0.00627 &-& 0.00557$^*$ \\ 
			& 2$p_{1/2}$ &	0.07935 & 0.24575 &	-&	-&	 &	0.00497 &	-&	- \\ 
			& 2$p_{3/2}$ &	0.07930 & 0.24431 &	-&	-&	&	0.00534 &	-&	- \\ 
			& 3$s$ & 0.07272 & 0.21199 & - & -&	\hspace*{-0.3cm}\rdelim\}{5}{*}[~0.032 (10)] &	0.01775 & -& 0.01614$^*$ \\ 
			& 3$p_{1/2}$ & 0.07265 & 0.21351 & - & - &	 &	0.01664 &	-&	- \\ 
			& 3$p_{3/2}$  & 0.07238 & 0.21190 &	-&	-&	&	0.01736 &	-&	- \\ 
			& 3$d_{3/2}$ &	0.07225 & 0.21613 &	-&	-	& &	0.01507 &	-&	- \\
			& 3$d_{5/2}$ & 0.07223 & 0.21556 &	-	&-& &	0.01532 &-	&- \\ 
			& 4$s$ & 0.06402 & 0.17603 &-	&-&	-	& 0.04148 &	-& 0.04291$^*$ \\  
			& 4$p_{1/2}$ &	0.06381 &	0.17462 &	-	&-	&-&	0.04226 &	-	&- \\
			& 4$p_{3/2}$ &  0.06343 &	0.17241 &-	&-	&-&	0.04387 &	-	&- \\
			& 4$d_{3/2}$ &	0.06333 &	0.16826 &-	&-	&-&	0.04640 &	-	&- \\
			& 4$d_{5/2}$ &	0.06314 &	0.16744 &-	& -& -& 0.04693 &	-&	- \\
			& 5$s$ &	0.05550 &	0.11864 &	-	&-	&-&	0.09147 &	-	&- \\
			& 5$p_{1/2}$ &	0.05438 &	0.10826 &	-&	-	&-&	0.09974 &-	&- \\
			& 5$p_{3/2}$ &	0.05437 &	0.10351 &	-&	-&	-&	0.10391	&-&	- \\ 
			& 5$d_{3/2}$ & - &	0.07259 & -	&-	&- & 0.15132 &- 	&-	 \\ 
			& 6$s$ &	0.04488 &	0.03833 &	-&	-&	-&	0.21868 &	-&	- \\  \hline \hline
			\multicolumn{5}{r}{$^{\dagger}$ With correlations.} & \multicolumn{5}{l}{$^*$ Interpolated.}  \\
	\end{longtable}

\clearpage

	\bibliography{ipcms}

\providecommand{\noopsort}[1]{}\providecommand{\singleletter}[1]{#1}%
\begin{thebibliography}{76}%
\makeatletter
\providecommand \@ifxundefined [1]{%
 \@ifx{#1\undefined}
}%
\providecommand \@ifnum [1]{%
 \ifnum #1\expandafter \@firstoftwo
 \else \expandafter \@secondoftwo
 \fi
}%
\providecommand \@ifx [1]{%
 \ifx #1\expandafter \@firstoftwo
 \else \expandafter \@secondoftwo
 \fi
}%
\providecommand \natexlab [1]{#1}%
\providecommand \enquote  [1]{``#1''}%
\providecommand \bibnamefont  [1]{#1}%
\providecommand \bibfnamefont [1]{#1}%
\providecommand \citenamefont [1]{#1}%
\providecommand \href@noop [0]{\@secondoftwo}%
\providecommand \href [0]{\begingroup \@sanitize@url \@href}%
\providecommand \@href[1]{\@@startlink{#1}\@@href}%
\providecommand \@@href[1]{\endgroup#1\@@endlink}%
\providecommand \@sanitize@url [0]{\catcode `\\12\catcode `\$12\catcode
  `\&12\catcode `\#12\catcode `\^12\catcode `\_12\catcode `\%12\relax}%
\providecommand \@@startlink[1]{}%
\providecommand \@@endlink[0]{}%
\providecommand \url  [0]{\begingroup\@sanitize@url \@url }%
\providecommand \@url [1]{\endgroup\@href {#1}{\urlprefix }}%
\providecommand \urlprefix  [0]{URL }%
\providecommand \Eprint [0]{\href }%
\providecommand \doibase [0]{http://dx.doi.org/}%
\providecommand \selectlanguage [0]{\@gobble}%
\providecommand \bibinfo  [0]{\@secondoftwo}%
\providecommand \bibfield  [0]{\@secondoftwo}%
\providecommand \translation [1]{[#1]}%
\providecommand \BibitemOpen [0]{}%
\providecommand \bibitemStop [0]{}%
\providecommand \bibitemNoStop [0]{.\EOS\space}%
\providecommand \EOS [0]{\spacefactor3000\relax}%
\providecommand \BibitemShut  [1]{\csname bibitem#1\endcsname}%
\let\auto@bib@innerbib\@empty
\bibitem [{\citenamefont {Coulon}\ \emph {et~al.}(2020)\citenamefont {Coulon},
  \citenamefont {Broda}, \citenamefont {Cassette}, \citenamefont {Courte},
  \citenamefont {Jerome}, \citenamefont {Judge}, \citenamefont {Kossert},
  \citenamefont {Liu}, \citenamefont {Michotte},\ and\ \citenamefont
  {Nonis}}]{Coulon20}%
  \BibitemOpen
  \bibfield  {author} {\bibinfo {author} {\bibfnamefont {R.}~\bibnamefont
  {Coulon}}, \bibinfo {author} {\bibfnamefont {R.}~\bibnamefont {Broda}},
  \bibinfo {author} {\bibfnamefont {P.}~\bibnamefont {Cassette}}, \bibinfo
  {author} {\bibfnamefont {S.}~\bibnamefont {Courte}}, \bibinfo {author}
  {\bibfnamefont {S.}~\bibnamefont {Jerome}}, \bibinfo {author} {\bibfnamefont
  {S.}~\bibnamefont {Judge}}, \bibinfo {author} {\bibfnamefont
  {K.}~\bibnamefont {Kossert}}, \bibinfo {author} {\bibfnamefont
  {H.}~\bibnamefont {Liu}}, \bibinfo {author} {\bibfnamefont {C.}~\bibnamefont
  {Michotte}}, \ and\ \bibinfo {author} {\bibfnamefont {M.}~\bibnamefont
  {Nonis}},\ }\href {\doibase 10.1088/1681-7575/ab7e7b} {\bibfield  {journal}
  {\bibinfo  {journal} {Metrologia}\ }\textbf {\bibinfo {volume} {57}},\
  \bibinfo {pages} {035009} (\bibinfo {year} {2020})}\BibitemShut {NoStop}%
\bibitem [{\citenamefont {Broda}\ \emph {et~al.}(2007)\citenamefont {Broda},
  \citenamefont {Cassette},\ and\ \citenamefont {Kossert}}]{Broda07}%
  \BibitemOpen
  \bibfield  {author} {\bibinfo {author} {\bibfnamefont {R.}~\bibnamefont
  {Broda}}, \bibinfo {author} {\bibfnamefont {P.}~\bibnamefont {Cassette}}, \
  and\ \bibinfo {author} {\bibfnamefont {K.}~\bibnamefont {Kossert}},\ }\href
  {\doibase 10.1088/0026-1394/44/4/s06} {\bibfield  {journal} {\bibinfo
  {journal} {Metrologia}\ }\textbf {\bibinfo {volume} {44}},\ \bibinfo {pages}
  {S36} (\bibinfo {year} {2007})}\BibitemShut {NoStop}%
\bibitem [{\citenamefont {Bavelaar}\ \emph {et~al.}(2018)\citenamefont
  {Bavelaar}, \citenamefont {Lee}, \citenamefont {Gill}, \citenamefont
  {Falzone},\ and\ \citenamefont {Vallis}}]{Bave18}%
  \BibitemOpen
  \bibfield  {author} {\bibinfo {author} {\bibfnamefont {B.~M.}\ \bibnamefont
  {Bavelaar}}, \bibinfo {author} {\bibfnamefont {B.~Q.}\ \bibnamefont {Lee}},
  \bibinfo {author} {\bibfnamefont {M.~R.}\ \bibnamefont {Gill}}, \bibinfo
  {author} {\bibfnamefont {N.}~\bibnamefont {Falzone}}, \ and\ \bibinfo
  {author} {\bibfnamefont {K.~A.}\ \bibnamefont {Vallis}},\ }\href {\doibase
  10.3389/fphar.2018.00996} {\bibfield  {journal} {\bibinfo  {journal}
  {Frontiers in Pharmacology}\ }\textbf {\bibinfo {volume} {9}},\ \bibinfo
  {pages} {996} (\bibinfo {year} {2018})}\BibitemShut {NoStop}%
\bibitem [{\citenamefont {Ku}\ \emph {et~al.}(2019)\citenamefont {Ku},
  \citenamefont {Facca}, \citenamefont {Cai},\ and\ \citenamefont
  {Reilly}}]{Ku19}%
  \BibitemOpen
  \bibfield  {author} {\bibinfo {author} {\bibfnamefont {A.}~\bibnamefont
  {Ku}}, \bibinfo {author} {\bibfnamefont {V.~J.}\ \bibnamefont {Facca}},
  \bibinfo {author} {\bibfnamefont {Z.}~\bibnamefont {Cai}}, \ and\ \bibinfo
  {author} {\bibfnamefont {R.~M.}\ \bibnamefont {Reilly}},\ }\href {\doibase
  10.1186/s41181-019-0075-2} {\bibfield  {journal} {\bibinfo  {journal} {EJNMMI
  Radiopharmacy and Chemistry}\ }\textbf {\bibinfo {volume} {4}},\ \bibinfo
  {pages} {27} (\bibinfo {year} {2019})}\BibitemShut {NoStop}%
\bibitem [{\citenamefont {Hou}(2005)}]{Hou05}%
  \BibitemOpen
  \bibfield  {author} {\bibinfo {author} {\bibfnamefont {X.}~\bibnamefont
  {Hou}},\ }\href {\doibase 10.1524/ract.2005.93.9-10.611} {\bibfield
  {journal} {\bibinfo  {journal} {Radiochimica Acta}\ }\textbf {\bibinfo
  {volume} {93}},\ \bibinfo {pages} {611} (\bibinfo {year} {2005})}\BibitemShut
  {NoStop}%
\bibitem [{\citenamefont {Lee}\ \emph {et~al.}(2021)\citenamefont {Lee},
  \citenamefont {Lim}, \citenamefont {Lee}, \citenamefont {Hong},\ and\
  \citenamefont {Kim}}]{Lee21}%
  \BibitemOpen
  \bibfield  {author} {\bibinfo {author} {\bibfnamefont {Y.-J.}\ \bibnamefont
  {Lee}}, \bibinfo {author} {\bibfnamefont {J.-M.}\ \bibnamefont {Lim}},
  \bibinfo {author} {\bibfnamefont {J.-H.}\ \bibnamefont {Lee}}, \bibinfo
  {author} {\bibfnamefont {S.-B.}\ \bibnamefont {Hong}}, \ and\ \bibinfo
  {author} {\bibfnamefont {H.}~\bibnamefont {Kim}},\ }\href {\doibase
  10.1016/j.net.2020.09.020} {\bibfield  {journal} {\bibinfo  {journal}
  {Nuclear Engineering and Technology}\ }\textbf {\bibinfo {volume} {53}},\
  \bibinfo {pages} {1210} (\bibinfo {year} {2021})}\BibitemShut {NoStop}%
\bibitem [{\citenamefont {Huss}\ \emph {et~al.}(2009)\citenamefont {Huss},
  \citenamefont {Meyer}, \citenamefont {Srinivasan}, \citenamefont {Goswami},\
  and\ \citenamefont {Sahijpal}}]{Huss09}%
  \BibitemOpen
  \bibfield  {author} {\bibinfo {author} {\bibfnamefont {G.~R.}\ \bibnamefont
  {Huss}}, \bibinfo {author} {\bibfnamefont {B.~S.}\ \bibnamefont {Meyer}},
  \bibinfo {author} {\bibfnamefont {G.}~\bibnamefont {Srinivasan}}, \bibinfo
  {author} {\bibfnamefont {J.~N.}\ \bibnamefont {Goswami}}, \ and\ \bibinfo
  {author} {\bibfnamefont {S.}~\bibnamefont {Sahijpal}},\ }\href {\doibase
  10.1016/j.gca.2009.01.039} {\bibfield  {journal} {\bibinfo  {journal}
  {Geochimica et Cosmochimica Acta}\ }\textbf {\bibinfo {volume} {73}},\
  \bibinfo {pages} {4922} (\bibinfo {year} {2009})}\BibitemShut {NoStop}%
\bibitem [{\citenamefont {Jörg}\ \emph {et~al.}(2012)\citenamefont {Jörg},
  \citenamefont {Amelin}, \citenamefont {Kossert},\ and\ \citenamefont {{Lierse
  v. Gostomski}}}]{Jorg12}%
  \BibitemOpen
  \bibfield  {author} {\bibinfo {author} {\bibfnamefont {G.}~\bibnamefont
  {Jörg}}, \bibinfo {author} {\bibfnamefont {Y.}~\bibnamefont {Amelin}},
  \bibinfo {author} {\bibfnamefont {K.}~\bibnamefont {Kossert}}, \ and\
  \bibinfo {author} {\bibfnamefont {C.}~\bibnamefont {{Lierse v. Gostomski}}},\
  }\href {\doibase 10.1016/j.gca.2012.03.036} {\bibfield  {journal} {\bibinfo
  {journal} {Geochimica et Cosmochimica Acta}\ }\textbf {\bibinfo {volume}
  {88}},\ \bibinfo {pages} {51} (\bibinfo {year} {2012})}\BibitemShut {NoStop}%
\bibitem [{\citenamefont {Kubik}\ \emph {et~al.}(1986)\citenamefont {Kubik},
  \citenamefont {Elmore}, \citenamefont {Conard}, \citenamefont {Nishiizumi},\
  and\ \citenamefont {Arnold}}]{Kubik86}%
  \BibitemOpen
  \bibfield  {author} {\bibinfo {author} {\bibfnamefont {P.~W.}\ \bibnamefont
  {Kubik}}, \bibinfo {author} {\bibfnamefont {D.}~\bibnamefont {Elmore}},
  \bibinfo {author} {\bibfnamefont {N.~J.}\ \bibnamefont {Conard}}, \bibinfo
  {author} {\bibfnamefont {K.}~\bibnamefont {Nishiizumi}}, \ and\ \bibinfo
  {author} {\bibfnamefont {J.~R.}\ \bibnamefont {Arnold}},\ }\href {\doibase
  10.1038/319568a0} {\bibfield  {journal} {\bibinfo  {journal} {Nature}\
  }\textbf {\bibinfo {volume} {319}},\ \bibinfo {pages} {568} (\bibinfo {year}
  {1986})}\BibitemShut {NoStop}%
\bibitem [{\citenamefont {Bhat}(1992)}]{ENSDF92}%
  \BibitemOpen
  \bibfield  {author} {\bibinfo {author} {\bibfnamefont {M.~R.}\ \bibnamefont
  {Bhat}},\ }in\ \href {\doibase 10.1007/978-3-642-58113-7_227} {\emph
  {\bibinfo {booktitle} {Nuclear Data for Science and Technology}}},\ \bibinfo
  {editor} {edited by\ \bibinfo {editor} {\bibfnamefont {S.~M.}\ \bibnamefont
  {Qaim}}}\ (\bibinfo  {publisher} {Springer},\ \bibinfo {address} {Berlin,
  Heidelberg, Germany},\ \bibinfo {year} {1992})\ pp.\ \bibinfo {pages}
  {817--821}\BibitemShut {NoStop}%
\bibitem [{\citenamefont {{Kellett, Mark A.}}\ and\ \citenamefont {{Bersillon,
  Olivier}}(2017)}]{DDEP17}%
  \BibitemOpen
  \bibfield  {author} {\bibinfo {author} {\bibnamefont {{Kellett, Mark A.}}}\
  and\ \bibinfo {author} {\bibnamefont {{Bersillon, Olivier}}},\ }\href
  {\doibase 10.1051/epjconf/201714602009} {\bibfield  {journal} {\bibinfo
  {journal} {EPJ Web Conf.}\ }\textbf {\bibinfo {volume} {146}},\ \bibinfo
  {pages} {02009} (\bibinfo {year} {2017})}\BibitemShut {NoStop}%
\bibitem [{\citenamefont {Gove}\ and\ \citenamefont {Martin}(1971)}]{Gove71}%
  \BibitemOpen
  \bibfield  {author} {\bibinfo {author} {\bibfnamefont {N.}~\bibnamefont
  {Gove}}\ and\ \bibinfo {author} {\bibfnamefont {M.}~\bibnamefont {Martin}},\
  }\href {\doibase 10.1016/S0092-640X(71)80026-8} {\bibfield  {journal}
  {\bibinfo  {journal} {Atomic Data and Nuclear Data Tables}\ }\textbf
  {\bibinfo {volume} {10}},\ \bibinfo {pages} {205} (\bibinfo {year}
  {1971})}\BibitemShut {NoStop}%
\bibitem [{\citenamefont {Mougeot}(2015)}]{Mougeot15}%
  \BibitemOpen
  \bibfield  {author} {\bibinfo {author} {\bibfnamefont {X.}~\bibnamefont
  {Mougeot}},\ }\href {\doibase 10.1103/PhysRevC.91.055504} {\bibfield
  {journal} {\bibinfo  {journal} {Phys. Rev. C}\ }\textbf {\bibinfo {volume}
  {91}},\ \bibinfo {pages} {055504} (\bibinfo {year} {2015})}\BibitemShut
  {NoStop}%
\bibitem [{\citenamefont {Mougeot}(2019)}]{Mougeot19}%
  \BibitemOpen
  \bibfield  {author} {\bibinfo {author} {\bibfnamefont {X.}~\bibnamefont
  {Mougeot}},\ }\href {\doibase 10.1016/j.apradiso.2019.108884} {\bibfield
  {journal} {\bibinfo  {journal} {Appl. Radiat. Isot.}\ }\textbf {\bibinfo
  {volume} {154}},\ \bibinfo {pages} {108884} (\bibinfo {year}
  {2019})}\BibitemShut {NoStop}%
\bibitem [{\citenamefont {Mougeot}(2018)}]{Mougeot18}%
  \BibitemOpen
  \bibfield  {author} {\bibinfo {author} {\bibfnamefont {X.}~\bibnamefont
  {Mougeot}},\ }\href {\doibase 10.1016/j.apradiso.2017.07.027} {\bibfield
  {journal} {\bibinfo  {journal} {Appl. Radiat. Isot.}\ }\textbf {\bibinfo
  {volume} {134}},\ \bibinfo {pages} {225} (\bibinfo {year}
  {2018})}\BibitemShut {NoStop}%
\bibitem [{\citenamefont {Ranitzsch}\ \emph {et~al.}(2020)\citenamefont
  {Ranitzsch}, \citenamefont {Arnold}, \citenamefont {Beyer}, \citenamefont
  {Bockhorn}, \citenamefont {Bonaparte}, \citenamefont {Enss}, \citenamefont
  {Kossert}, \citenamefont {Kempf}, \citenamefont {Loidl}, \citenamefont
  {Mariam}, \citenamefont {N{\"a}hle}, \citenamefont {Paulsen}, \citenamefont
  {Rodrigues},\ and\ \citenamefont {Wegner}}]{MetroMMC20}%
  \BibitemOpen
  \bibfield  {author} {\bibinfo {author} {\bibfnamefont {P.~C.-O.}\
  \bibnamefont {Ranitzsch}}, \bibinfo {author} {\bibfnamefont {D.}~\bibnamefont
  {Arnold}}, \bibinfo {author} {\bibfnamefont {J.}~\bibnamefont {Beyer}},
  \bibinfo {author} {\bibfnamefont {L.}~\bibnamefont {Bockhorn}}, \bibinfo
  {author} {\bibfnamefont {J.~J.}\ \bibnamefont {Bonaparte}}, \bibinfo {author}
  {\bibfnamefont {C.}~\bibnamefont {Enss}}, \bibinfo {author} {\bibfnamefont
  {K.}~\bibnamefont {Kossert}}, \bibinfo {author} {\bibfnamefont
  {S.}~\bibnamefont {Kempf}}, \bibinfo {author} {\bibfnamefont
  {M.}~\bibnamefont {Loidl}}, \bibinfo {author} {\bibfnamefont
  {R.}~\bibnamefont {Mariam}}, \bibinfo {author} {\bibfnamefont {O.~J.}\
  \bibnamefont {N{\"a}hle}}, \bibinfo {author} {\bibfnamefont {M.}~\bibnamefont
  {Paulsen}}, \bibinfo {author} {\bibfnamefont {M.}~\bibnamefont {Rodrigues}},
  \ and\ \bibinfo {author} {\bibfnamefont {M.}~\bibnamefont {Wegner}},\ }\href
  {\doibase 10.1007/s10909-019-02278-4} {\bibfield  {journal} {\bibinfo
  {journal} {Journal of Low Temperature Physics}\ }\textbf {\bibinfo {volume}
  {199}},\ \bibinfo {pages} {441} (\bibinfo {year} {2020})}\BibitemShut
  {NoStop}%
\bibitem [{Pri(2024)}]{PrimaLTD}%
  \BibitemOpen
  \href@noop {} {\enquote {\bibinfo {title} {{The EMPIR PrimA-LTD Project}},}\
  }\bibinfo {howpublished} {\url{https://prima-ltd.net/}} (\bibinfo {year}
  {2021-2024})\BibitemShut {NoStop}%
\bibitem [{\citenamefont {Bambynek}\ \emph {et~al.}(1977)\citenamefont
  {Bambynek}, \citenamefont {Behrens}, \citenamefont {Chen}, \citenamefont
  {Crasemann}, \citenamefont {Fitzpatrick}, \citenamefont {Ledingham},
  \citenamefont {Genz}, \citenamefont {Mutterer},\ and\ \citenamefont
  {Intemann}}]{Bamb77}%
  \BibitemOpen
  \bibfield  {author} {\bibinfo {author} {\bibfnamefont {W.}~\bibnamefont
  {Bambynek}}, \bibinfo {author} {\bibfnamefont {H.}~\bibnamefont {Behrens}},
  \bibinfo {author} {\bibfnamefont {M.~H.}\ \bibnamefont {Chen}}, \bibinfo
  {author} {\bibfnamefont {B.}~\bibnamefont {Crasemann}}, \bibinfo {author}
  {\bibfnamefont {M.~L.}\ \bibnamefont {Fitzpatrick}}, \bibinfo {author}
  {\bibfnamefont {K.~W.~D.}\ \bibnamefont {Ledingham}}, \bibinfo {author}
  {\bibfnamefont {H.}~\bibnamefont {Genz}}, \bibinfo {author} {\bibfnamefont
  {M.}~\bibnamefont {Mutterer}}, \ and\ \bibinfo {author} {\bibfnamefont
  {R.~L.}\ \bibnamefont {Intemann}},\ }\href {\doibase
  10.1103/RevModPhys.49.77} {\bibfield  {journal} {\bibinfo  {journal} {Rev.
  Mod. Phys.}\ }\textbf {\bibinfo {volume} {49}},\ \bibinfo {pages} {77}
  (\bibinfo {year} {1977})}\BibitemShut {NoStop}%
\bibitem [{\citenamefont {Behrens}\ and\ \citenamefont
  {B{\"u}hring}(1982)}]{Behrens82}%
  \BibitemOpen
  \bibfield  {author} {\bibinfo {author} {\bibfnamefont {H.}~\bibnamefont
  {Behrens}}\ and\ \bibinfo {author} {\bibfnamefont {W.}~\bibnamefont
  {B{\"u}hring}},\ }\href@noop {} {\emph {\bibinfo {title} {Electron Radial
  Wave Functions and Nuclear Beta Decay}}}\ (\bibinfo  {publisher}
  {Clarendon},\ \bibinfo {address} {Oxford, UK},\ \bibinfo {year}
  {1982})\BibitemShut {NoStop}%
\bibitem [{\citenamefont {Bahcall}(1965)}]{Bahcall65}%
  \BibitemOpen
  \bibfield  {author} {\bibinfo {author} {\bibfnamefont {J.~N.}\ \bibnamefont
  {Bahcall}},\ }\href {\doibase 10.1016/0029-5582(65)90717-0} {\bibfield
  {journal} {\bibinfo  {journal} {Nuclear Physics}\ }\textbf {\bibinfo {volume}
  {71}},\ \bibinfo {pages} {267} (\bibinfo {year} {1965})}\BibitemShut
  {NoStop}%
\bibitem [{\citenamefont {Vatai}(1970)}]{Vatai70}%
  \BibitemOpen
  \bibfield  {author} {\bibinfo {author} {\bibfnamefont {E.}~\bibnamefont
  {Vatai}},\ }\href {\doibase 10.1016/0375-9474(70)90250-2} {\bibfield
  {journal} {\bibinfo  {journal} {Nuclear Physics A}\ }\textbf {\bibinfo
  {volume} {156}},\ \bibinfo {pages} {541} (\bibinfo {year}
  {1970})}\BibitemShut {NoStop}%
\bibitem [{\citenamefont {Crasemann}\ \emph {et~al.}(1979)\citenamefont
  {Crasemann}, \citenamefont {Chen}, \citenamefont {Briand}, \citenamefont
  {Chevallier}, \citenamefont {Chetioui},\ and\ \citenamefont
  {Tavernier}}]{Crase79}%
  \BibitemOpen
  \bibfield  {author} {\bibinfo {author} {\bibfnamefont {B.}~\bibnamefont
  {Crasemann}}, \bibinfo {author} {\bibfnamefont {M.~H.}\ \bibnamefont {Chen}},
  \bibinfo {author} {\bibfnamefont {J.~P.}\ \bibnamefont {Briand}}, \bibinfo
  {author} {\bibfnamefont {P.}~\bibnamefont {Chevallier}}, \bibinfo {author}
  {\bibfnamefont {A.}~\bibnamefont {Chetioui}}, \ and\ \bibinfo {author}
  {\bibfnamefont {M.}~\bibnamefont {Tavernier}},\ }\href {\doibase
  10.1103/PhysRevC.19.1042} {\bibfield  {journal} {\bibinfo  {journal} {Phys.
  Rev. C}\ }\textbf {\bibinfo {volume} {19}},\ \bibinfo {pages} {1042}
  (\bibinfo {year} {1979})}\BibitemShut {NoStop}%
\bibitem [{\citenamefont {Mougeot}\ and\ \citenamefont
  {Bisch}(2014)}]{Mougeot14}%
  \BibitemOpen
  \bibfield  {author} {\bibinfo {author} {\bibfnamefont {X.}~\bibnamefont
  {Mougeot}}\ and\ \bibinfo {author} {\bibfnamefont {C.}~\bibnamefont
  {Bisch}},\ }\href {\doibase 10.1103/PhysRevA.90.012501} {\bibfield  {journal}
  {\bibinfo  {journal} {Phys. Rev. A}\ }\textbf {\bibinfo {volume} {90}},\
  \bibinfo {pages} {012501} (\bibinfo {year} {2014})}\BibitemShut {NoStop}%
\bibitem [{\citenamefont {Salvat}\ \emph {et~al.}(1987)\citenamefont {Salvat},
  \citenamefont {Mart{\'i}nez}, \citenamefont {Mayol},\ and\ \citenamefont
  {Parellada}}]{Salvat87}%
  \BibitemOpen
  \bibfield  {author} {\bibinfo {author} {\bibfnamefont {F.}~\bibnamefont
  {Salvat}}, \bibinfo {author} {\bibfnamefont {J.~D.}\ \bibnamefont
  {Mart{\'i}nez}}, \bibinfo {author} {\bibfnamefont {R.}~\bibnamefont {Mayol}},
  \ and\ \bibinfo {author} {\bibfnamefont {J.}~\bibnamefont {Parellada}},\
  }\href {\doibase 10.1103/PhysRevA.36.467} {\bibfield  {journal} {\bibinfo
  {journal} {Phys. Rev. A}\ }\textbf {\bibinfo {volume} {36}},\ \bibinfo
  {pages} {467} (\bibinfo {year} {1987})}\BibitemShut {NoStop}%
\bibitem [{\citenamefont {Desclaux}(1973)}]{Desclaux73}%
  \BibitemOpen
  \bibfield  {author} {\bibinfo {author} {\bibfnamefont {J.}~\bibnamefont
  {Desclaux}},\ }\href {\doibase 10.1016/0092-640X(73)90020-X} {\bibfield
  {journal} {\bibinfo  {journal} {Atomic Data and Nuclear Data Tables}\
  }\textbf {\bibinfo {volume} {12}},\ \bibinfo {pages} {311} (\bibinfo {year}
  {1973})}\BibitemShut {NoStop}%
\bibitem [{\citenamefont {Kotochigova}\ \emph {et~al.}(1997)\citenamefont
  {Kotochigova}, \citenamefont {Levine}, \citenamefont {Shirley}, \citenamefont
  {Stiles},\ and\ \citenamefont {Clark}}]{Kot97}%
  \BibitemOpen
  \bibfield  {author} {\bibinfo {author} {\bibfnamefont {S.}~\bibnamefont
  {Kotochigova}}, \bibinfo {author} {\bibfnamefont {Z.~H.}\ \bibnamefont
  {Levine}}, \bibinfo {author} {\bibfnamefont {E.~L.}\ \bibnamefont {Shirley}},
  \bibinfo {author} {\bibfnamefont {M.~D.}\ \bibnamefont {Stiles}}, \ and\
  \bibinfo {author} {\bibfnamefont {C.~W.}\ \bibnamefont {Clark}},\ }\href
  {\doibase 10.1103/PhysRevA.55.191} {\bibfield  {journal} {\bibinfo  {journal}
  {Phys. Rev. A}\ }\textbf {\bibinfo {volume} {55}},\ \bibinfo {pages} {191}
  (\bibinfo {year} {1997})}\BibitemShut {NoStop}%
\bibitem [{\citenamefont {Kotochigova}\ \emph {et~al.}(2021)\citenamefont
  {Kotochigova}, \citenamefont {Levine}, \citenamefont {Shirley}, \citenamefont
  {Stiles},\ and\ \citenamefont {Clark}}]{NIST09}%
  \BibitemOpen
  \bibfield  {author} {\bibinfo {author} {\bibfnamefont {S.}~\bibnamefont
  {Kotochigova}}, \bibinfo {author} {\bibfnamefont {Z.~H.}\ \bibnamefont
  {Levine}}, \bibinfo {author} {\bibfnamefont {E.~L.}\ \bibnamefont {Shirley}},
  \bibinfo {author} {\bibfnamefont {M.~D.}\ \bibnamefont {Stiles}}, \ and\
  \bibinfo {author} {\bibfnamefont {C.~W.}\ \bibnamefont {Clark}},\ }\href
  {\doibase 10.18434/T4ZP4F} {\emph {\bibinfo {title} {NIST Chemistry WebBook,
  NIST Standard Reference Database 141}}},\ edited by\ \bibinfo {editor}
  {\bibfnamefont {P.}~\bibnamefont {Linstrom}}\ and\ \bibinfo {editor}
  {\bibfnamefont {W.}~\bibnamefont {Mallard}}\ (\bibinfo  {publisher} {National
  Institute of Standards and Technology},\ \bibinfo {address} {Gaithersburg,
  MA, 20899},\ \bibinfo {year} {2021})\ Chap.\ \bibinfo {chapter} {Atomic
  Reference Data for Electronic Structure Calculations}\BibitemShut {NoStop}%
\bibitem [{\citenamefont {Desclaux}(1975)}]{DESCLAUX1975}%
  \BibitemOpen
  \bibfield  {author} {\bibinfo {author} {\bibfnamefont {J.}~\bibnamefont
  {Desclaux}},\ }\href {\doibase https://doi.org/10.1016/0010-4655(75)90054-5}
  {\bibfield  {journal} {\bibinfo  {journal} {Computer Physics Communications}\
  }\textbf {\bibinfo {volume} {9}},\ \bibinfo {pages} {31} (\bibinfo {year}
  {1975})}\BibitemShut {NoStop}%
\bibitem [{\citenamefont {Indelicato}\ and\ \citenamefont
  {Desclaux}(1990)}]{INDELICATO1990}%
  \BibitemOpen
  \bibfield  {author} {\bibinfo {author} {\bibfnamefont {P.}~\bibnamefont
  {Indelicato}}\ and\ \bibinfo {author} {\bibfnamefont {J.~P.}\ \bibnamefont
  {Desclaux}},\ }\href {\doibase 10.1103/PhysRevA.42.5139} {\bibfield
  {journal} {\bibinfo  {journal} {Phys. Rev. A}\ }\textbf {\bibinfo {volume}
  {42}},\ \bibinfo {pages} {5139} (\bibinfo {year} {1990})}\BibitemShut
  {NoStop}%
\bibitem [{\citenamefont {P{\'a}lffy}(2010)}]{Palffy10}%
  \BibitemOpen
  \bibfield  {author} {\bibinfo {author} {\bibfnamefont {A.}~\bibnamefont
  {P{\'a}lffy}},\ }\href {\doibase 10.1080/00107514.2010.493325} {\bibfield
  {journal} {\bibinfo  {journal} {Contemporary Physics}\ }\textbf {\bibinfo
  {volume} {51}},\ \bibinfo {pages} {471} (\bibinfo {year} {2010})}\BibitemShut
  {NoStop}%
\bibitem [{\citenamefont {Audi}\ \emph {et~al.}(2003)\citenamefont {Audi},
  \citenamefont {Wapstra},\ and\ \citenamefont {Thibault}}]{AUDI03}%
  \BibitemOpen
  \bibfield  {author} {\bibinfo {author} {\bibfnamefont {G.}~\bibnamefont
  {Audi}}, \bibinfo {author} {\bibfnamefont {A.}~\bibnamefont {Wapstra}}, \
  and\ \bibinfo {author} {\bibfnamefont {C.}~\bibnamefont {Thibault}},\ }\href
  {\doibase https://doi.org/10.1016/j.nuclphysa.2003.11.003} {\bibfield
  {journal} {\bibinfo  {journal} {Nuclear Physics A}\ }\textbf {\bibinfo
  {volume} {729}},\ \bibinfo {pages} {337} (\bibinfo {year} {2003})},\ \bibinfo
  {note} {the 2003 NUBASE and Atomic Mass Evaluations}\BibitemShut {NoStop}%
\bibitem [{\citenamefont {Angeli}\ and\ \citenamefont
  {Marinova}(2013)}]{ANGELI2013}%
  \BibitemOpen
  \bibfield  {author} {\bibinfo {author} {\bibfnamefont {I.}~\bibnamefont
  {Angeli}}\ and\ \bibinfo {author} {\bibfnamefont {K.}~\bibnamefont
  {Marinova}},\ }\href {\doibase https://doi.org/10.1016/j.adt.2011.12.006}
  {\bibfield  {journal} {\bibinfo  {journal} {Atomic Data and Nuclear Data
  Tables}\ }\textbf {\bibinfo {volume} {99}},\ \bibinfo {pages} {69} (\bibinfo
  {year} {2013})}\BibitemShut {NoStop}%
\bibitem [{\citenamefont {Grant}\ and\ \citenamefont
  {Quiney}(1988)}]{GRANT198837}%
  \BibitemOpen
  \bibfield  {author} {\bibinfo {author} {\bibfnamefont {I.~P.}\ \bibnamefont
  {Grant}}\ and\ \bibinfo {author} {\bibfnamefont {H.~M.}\ \bibnamefont
  {Quiney}},\ }\href {\doibase 10.1016/S0065-2199(08)60105-0} {\bibfield
  {journal} {\bibinfo  {journal} {Adv. Atom. Mol. Phys.}\ }\textbf {\bibinfo
  {volume} {23}},\ \bibinfo {pages} {37} (\bibinfo {year} {1988})}\BibitemShut
  {NoStop}%
\bibitem [{\citenamefont {Indelicato}(1995)}]{INDELICATO1995}%
  \BibitemOpen
  \bibfield  {author} {\bibinfo {author} {\bibfnamefont {P.}~\bibnamefont
  {Indelicato}},\ }\href {\doibase 10.1103/PhysRevA.51.1132} {\bibfield
  {journal} {\bibinfo  {journal} {Phys. Rev. A}\ }\textbf {\bibinfo {volume}
  {51}},\ \bibinfo {pages} {1132} (\bibinfo {year} {1995})}\BibitemShut
  {NoStop}%
\bibitem [{\citenamefont {Parpia}\ \emph {et~al.}(1996)\citenamefont {Parpia},
  \citenamefont {Fischer},\ and\ \citenamefont {Grant}}]{PARPIA1996}%
  \BibitemOpen
  \bibfield  {author} {\bibinfo {author} {\bibfnamefont {F.}~\bibnamefont
  {Parpia}}, \bibinfo {author} {\bibfnamefont {C.}~\bibnamefont {Fischer}}, \
  and\ \bibinfo {author} {\bibfnamefont {I.}~\bibnamefont {Grant}},\ }\href
  {\doibase https://doi.org/10.1016/0010-4655(95)00136-0} {\bibfield  {journal}
  {\bibinfo  {journal} {Computer Physics Communications}\ }\textbf {\bibinfo
  {volume} {94}},\ \bibinfo {pages} {249} (\bibinfo {year} {1996})}\BibitemShut
  {NoStop}%
\bibitem [{\citenamefont {Dreizler}\ and\ \citenamefont
  {Engel}(1998)}]{Dreizler1998}%
  \BibitemOpen
  \bibfield  {author} {\bibinfo {author} {\bibfnamefont {R.~M.}\ \bibnamefont
  {Dreizler}}\ and\ \bibinfo {author} {\bibfnamefont {E.}~\bibnamefont
  {Engel}},\ }\enquote {\bibinfo {title} {Relativistic density functional
  theory},}\ in\ \href {\doibase 10.1007/BFb0106736} {\emph {\bibinfo
  {booktitle} {Density Functionals: Theory and Applications: Proceedings of the
  Tenth Chris Engelbrecht Summer School in Theoretical Physics Held at
  Meerensee, near Cape Town South Africa, 19--29 January 1997}}},\ \bibinfo
  {editor} {edited by\ \bibinfo {editor} {\bibfnamefont {D.}~\bibnamefont
  {Joubert}}}\ (\bibinfo  {publisher} {Springer},\ \bibinfo {address} {Berlin,
  Heidelberg, Germany},\ \bibinfo {year} {1998})\ pp.\ \bibinfo {pages}
  {147--189}\BibitemShut {NoStop}%
\bibitem [{\citenamefont {Rajagopal}(1978)}]{Rajagopal_1978}%
  \BibitemOpen
  \bibfield  {author} {\bibinfo {author} {\bibfnamefont {A.~K.}\ \bibnamefont
  {Rajagopal}},\ }\href {\doibase 10.1088/0022-3719/11/24/002} {\bibfield
  {journal} {\bibinfo  {journal} {Journal of Physics C: Solid State Physics}\
  }\textbf {\bibinfo {volume} {11}},\ \bibinfo {pages} {L943} (\bibinfo {year}
  {1978})}\BibitemShut {NoStop}%
\bibitem [{\citenamefont {MacDonald}\ and\ \citenamefont
  {Vosko}(1979)}]{MacDonald_1979}%
  \BibitemOpen
  \bibfield  {author} {\bibinfo {author} {\bibfnamefont {A.~H.}\ \bibnamefont
  {MacDonald}}\ and\ \bibinfo {author} {\bibfnamefont {S.~H.}\ \bibnamefont
  {Vosko}},\ }\href {\doibase 10.1088/0022-3719/12/15/007} {\bibfield
  {journal} {\bibinfo  {journal} {Journal of Physics C: Solid State Physics}\
  }\textbf {\bibinfo {volume} {12}},\ \bibinfo {pages} {2977} (\bibinfo {year}
  {1979})}\BibitemShut {NoStop}%
\bibitem [{\citenamefont {Engel}(2002)}]{Engel02}%
  \BibitemOpen
  \bibfield  {author} {\bibinfo {author} {\bibfnamefont {E.}~\bibnamefont
  {Engel}},\ }\href {https://itp.uni-frankfurt.de/~engel/papers/e02.pdf} {\emph
  {\bibinfo {title} {Relativistic Electronic Structure Theory, Part 1.
  Fundamentals}}},\ edited by\ \bibinfo {editor} {\bibfnamefont
  {P.}~\bibnamefont {Schwerdtfeger}}\ (\bibinfo  {publisher} {Elsevier},\
  \bibinfo {address} {Amsterdam, the Netherlands},\ \bibinfo {year} {2002})\
  pp.\ \bibinfo {pages} {524--624}\BibitemShut {NoStop}%
\bibitem [{\citenamefont {Strange}(2008)}]{Strange08}%
  \BibitemOpen
  \bibfield  {author} {\bibinfo {author} {\bibfnamefont {P.}~\bibnamefont
  {Strange}},\ }\href@noop {} {\emph {\bibinfo {title} {Relativistic Quantum
  Mechanics: With Applications in Condensed Matter and Atomic Physics}}}\
  (\bibinfo  {publisher} {Cambridge University Press},\ \bibinfo {address}
  {Cambridge, UK},\ \bibinfo {year} {2008})\BibitemShut {NoStop}%
\bibitem [{\citenamefont {Johnson}\ and\ \citenamefont
  {Soff}(1985)}]{JOHNSON1985405}%
  \BibitemOpen
  \bibfield  {author} {\bibinfo {author} {\bibfnamefont {W.}~\bibnamefont
  {Johnson}}\ and\ \bibinfo {author} {\bibfnamefont {G.}~\bibnamefont {Soff}},\
  }\href {\doibase 10.1016/0092-640X(85)90010-5} {\bibfield  {journal}
  {\bibinfo  {journal} {Atomic Data and Nuclear Data Tables}\ }\textbf
  {\bibinfo {volume} {33}},\ \bibinfo {pages} {405} (\bibinfo {year}
  {1985})}\BibitemShut {NoStop}%
\bibitem [{\citenamefont {Vosko}\ \emph {et~al.}(1980)\citenamefont {Vosko},
  \citenamefont {Wilk},\ and\ \citenamefont {Nusair}}]{Vosko80}%
  \BibitemOpen
  \bibfield  {author} {\bibinfo {author} {\bibfnamefont {S.~H.}\ \bibnamefont
  {Vosko}}, \bibinfo {author} {\bibfnamefont {L.}~\bibnamefont {Wilk}}, \ and\
  \bibinfo {author} {\bibfnamefont {M.}~\bibnamefont {Nusair}},\ }\href
  {\doibase 10.1139/p80-159} {\bibfield  {journal} {\bibinfo  {journal}
  {Canadian Journal of Physics}\ }\textbf {\bibinfo {volume} {58}},\ \bibinfo
  {pages} {1200} (\bibinfo {year} {1980})}\BibitemShut {NoStop}%
\bibitem [{\citenamefont {Gunnarsson}\ and\ \citenamefont
  {Lundqvist}(1976)}]{Gun76}%
  \BibitemOpen
  \bibfield  {author} {\bibinfo {author} {\bibfnamefont {O.}~\bibnamefont
  {Gunnarsson}}\ and\ \bibinfo {author} {\bibfnamefont {B.~I.}\ \bibnamefont
  {Lundqvist}},\ }\href {\doibase 10.1103/PhysRevB.13.4274} {\bibfield
  {journal} {\bibinfo  {journal} {Phys. Rev. B}\ }\textbf {\bibinfo {volume}
  {13}},\ \bibinfo {pages} {4274} (\bibinfo {year} {1976})}\BibitemShut
  {NoStop}%
\bibitem [{\citenamefont {Perdew}\ and\ \citenamefont {Wang}(1992)}]{Perd92}%
  \BibitemOpen
  \bibfield  {author} {\bibinfo {author} {\bibfnamefont {J.~P.}\ \bibnamefont
  {Perdew}}\ and\ \bibinfo {author} {\bibfnamefont {Y.}~\bibnamefont {Wang}},\
  }\href {\doibase 10.1103/PhysRevB.45.13244} {\bibfield  {journal} {\bibinfo
  {journal} {Phys. Rev. B}\ }\textbf {\bibinfo {volume} {45}},\ \bibinfo
  {pages} {13244} (\bibinfo {year} {1992})}\BibitemShut {NoStop}%
\bibitem [{\citenamefont {van Leeuwen}\ and\ \citenamefont
  {Baerends}(1994)}]{Leeuwen94}%
  \BibitemOpen
  \bibfield  {author} {\bibinfo {author} {\bibfnamefont {R.}~\bibnamefont {van
  Leeuwen}}\ and\ \bibinfo {author} {\bibfnamefont {E.~J.}\ \bibnamefont
  {Baerends}},\ }\href {\doibase 10.1103/PhysRevA.49.2421} {\bibfield
  {journal} {\bibinfo  {journal} {Phys. Rev. A}\ }\textbf {\bibinfo {volume}
  {49}},\ \bibinfo {pages} {2421} (\bibinfo {year} {1994})}\BibitemShut
  {NoStop}%
\bibitem [{\citenamefont {{\v{C}}ert{\'i}k}\ \emph {et~al.}(2013)\citenamefont
  {{\v{C}}ert{\'i}k}, \citenamefont {Pask},\ and\ \citenamefont
  {Vack{\'a}{\v{r}}}}]{CERTIK20131777}%
  \BibitemOpen
  \bibfield  {author} {\bibinfo {author} {\bibfnamefont {O.}~\bibnamefont
  {{\v{C}}ert{\'i}k}}, \bibinfo {author} {\bibfnamefont {J.~E.}\ \bibnamefont
  {Pask}}, \ and\ \bibinfo {author} {\bibfnamefont {J.}~\bibnamefont
  {Vack{\'a}{\v{r}}}},\ }\href {\doibase 10.1016/j.cpc.2013.02.014} {\bibfield
  {journal} {\bibinfo  {journal} {Computer Physics Communications}\ }\textbf
  {\bibinfo {volume} {184}},\ \bibinfo {pages} {1777} (\bibinfo {year}
  {2013})}\BibitemShut {NoStop}%
\bibitem [{\citenamefont {Ciofini}\ \emph {et~al.}(2005)\citenamefont
  {Ciofini}, \citenamefont {Adamo},\ and\ \citenamefont
  {Chermette}}]{CIOFINI200567}%
  \BibitemOpen
  \bibfield  {author} {\bibinfo {author} {\bibfnamefont {I.}~\bibnamefont
  {Ciofini}}, \bibinfo {author} {\bibfnamefont {C.}~\bibnamefont {Adamo}}, \
  and\ \bibinfo {author} {\bibfnamefont {H.}~\bibnamefont {Chermette}},\ }\href
  {\doibase 10.1016/j.chemphys.2004.05.034} {\bibfield  {journal} {\bibinfo
  {journal} {Chemical Physics}\ }\textbf {\bibinfo {volume} {309}},\ \bibinfo
  {pages} {67} (\bibinfo {year} {2005})},\ \bibinfo {note} {electronic
  Structure Calculations for Understanding Surfaces and Molecules. In Honor of
  Notker Roesch}\BibitemShut {NoStop}%
\bibitem [{\citenamefont {Perdew}\ and\ \citenamefont
  {Zunger}(1981)}]{Perdew81}%
  \BibitemOpen
  \bibfield  {author} {\bibinfo {author} {\bibfnamefont {J.~P.}\ \bibnamefont
  {Perdew}}\ and\ \bibinfo {author} {\bibfnamefont {A.}~\bibnamefont
  {Zunger}},\ }\href {\doibase 10.1103/PhysRevB.23.5048} {\bibfield  {journal}
  {\bibinfo  {journal} {Phys. Rev. B}\ }\textbf {\bibinfo {volume} {23}},\
  \bibinfo {pages} {5048} (\bibinfo {year} {1981})}\BibitemShut {NoStop}%
\bibitem [{\citenamefont {Politis}\ \emph {et~al.}(1998)\citenamefont
  {Politis}, \citenamefont {Hervieux}, \citenamefont {Hanssen}, \citenamefont
  {Madjet},\ and\ \citenamefont {Mart\'{\i}n}}]{Politis98}%
  \BibitemOpen
  \bibfield  {author} {\bibinfo {author} {\bibfnamefont {M.~F.}\ \bibnamefont
  {Politis}}, \bibinfo {author} {\bibfnamefont {P.~A.}\ \bibnamefont
  {Hervieux}}, \bibinfo {author} {\bibfnamefont {J.}~\bibnamefont {Hanssen}},
  \bibinfo {author} {\bibfnamefont {M.~E.}\ \bibnamefont {Madjet}}, \ and\
  \bibinfo {author} {\bibfnamefont {F.}~\bibnamefont {Mart\'{\i}n}},\ }\href
  {\doibase 10.1103/PhysRevA.58.367} {\bibfield  {journal} {\bibinfo  {journal}
  {Phys. Rev. A}\ }\textbf {\bibinfo {volume} {58}},\ \bibinfo {pages} {367}
  (\bibinfo {year} {1998})}\BibitemShut {NoStop}%
\bibitem [{\citenamefont {Krieger}\ \emph {et~al.}(1992)\citenamefont
  {Krieger}, \citenamefont {Li},\ and\ \citenamefont {Iafrate}}]{Kri92}%
  \BibitemOpen
  \bibfield  {author} {\bibinfo {author} {\bibfnamefont {J.~B.}\ \bibnamefont
  {Krieger}}, \bibinfo {author} {\bibfnamefont {Y.}~\bibnamefont {Li}}, \ and\
  \bibinfo {author} {\bibfnamefont {G.~J.}\ \bibnamefont {Iafrate}},\ }\href
  {\doibase 10.1103/PhysRevA.46.5453} {\bibfield  {journal} {\bibinfo
  {journal} {Phys. Rev. A}\ }\textbf {\bibinfo {volume} {46}},\ \bibinfo
  {pages} {5453} (\bibinfo {year} {1992})}\BibitemShut {NoStop}%
\bibitem [{\citenamefont {Li}\ \emph {et~al.}(1993)\citenamefont {Li},
  \citenamefont {Krieger},\ and\ \citenamefont {Iafrate}}]{Kri93}%
  \BibitemOpen
  \bibfield  {author} {\bibinfo {author} {\bibfnamefont {Y.}~\bibnamefont
  {Li}}, \bibinfo {author} {\bibfnamefont {J.~B.}\ \bibnamefont {Krieger}}, \
  and\ \bibinfo {author} {\bibfnamefont {G.~J.}\ \bibnamefont {Iafrate}},\
  }\href {\doibase 10.1103/PhysRevA.47.165} {\bibfield  {journal} {\bibinfo
  {journal} {Phys. Rev. A}\ }\textbf {\bibinfo {volume} {47}},\ \bibinfo
  {pages} {165} (\bibinfo {year} {1993})}\BibitemShut {NoStop}%
\bibitem [{\citenamefont {Tong}\ and\ \citenamefont {Chu}(1998)}]{Tong98}%
  \BibitemOpen
  \bibfield  {author} {\bibinfo {author} {\bibfnamefont {X.-M.}\ \bibnamefont
  {Tong}}\ and\ \bibinfo {author} {\bibfnamefont {S.-I.}\ \bibnamefont {Chu}},\
  }\href {\doibase 10.1103/PhysRevA.57.855} {\bibfield  {journal} {\bibinfo
  {journal} {Phys. Rev. A}\ }\textbf {\bibinfo {volume} {57}},\ \bibinfo
  {pages} {855} (\bibinfo {year} {1998})}\BibitemShut {NoStop}%
\bibitem [{\citenamefont {Sevier}(1979)}]{Sevier79}%
  \BibitemOpen
  \bibfield  {author} {\bibinfo {author} {\bibfnamefont {K.~D.}\ \bibnamefont
  {Sevier}},\ }\href {\doibase 10.1016/0092-640X(79)90012-3} {\bibfield
  {journal} {\bibinfo  {journal} {Atomic Data and Nuclear Data Tables}\
  }\textbf {\bibinfo {volume} {24}},\ \bibinfo {pages} {323} (\bibinfo {year}
  {1979})}\BibitemShut {NoStop}%
\bibitem [{\citenamefont {Bearden}\ and\ \citenamefont
  {Burr}(1967)}]{Bearden67}%
  \BibitemOpen
  \bibfield  {author} {\bibinfo {author} {\bibfnamefont {J.~A.}\ \bibnamefont
  {Bearden}}\ and\ \bibinfo {author} {\bibfnamefont {A.~F.}\ \bibnamefont
  {Burr}},\ }\href {\doibase 10.1103/RevModPhys.39.125} {\bibfield  {journal}
  {\bibinfo  {journal} {Rev. Mod. Phys.}\ }\textbf {\bibinfo {volume} {39}},\
  \bibinfo {pages} {125} (\bibinfo {year} {1967})}\BibitemShut {NoStop}%
\bibitem [{\citenamefont {Mukoyama}\ and\ \citenamefont
  {Taniguchi}(1987)}]{PhysRevA.36.693}%
  \BibitemOpen
  \bibfield  {author} {\bibinfo {author} {\bibfnamefont {T.}~\bibnamefont
  {Mukoyama}}\ and\ \bibinfo {author} {\bibfnamefont {K.}~\bibnamefont
  {Taniguchi}},\ }\href {\doibase 10.1103/PhysRevA.36.693} {\bibfield
  {journal} {\bibinfo  {journal} {Phys. Rev. A}\ }\textbf {\bibinfo {volume}
  {36}},\ \bibinfo {pages} {693} (\bibinfo {year} {1987})}\BibitemShut
  {NoStop}%
\bibitem [{\citenamefont {Kochur}\ \emph {et~al.}(2002)\citenamefont {Kochur},
  \citenamefont {Dudenko},\ and\ \citenamefont {Petrini}}]{AGKochur_2002}%
  \BibitemOpen
  \bibfield  {author} {\bibinfo {author} {\bibfnamefont {A.~G.}\ \bibnamefont
  {Kochur}}, \bibinfo {author} {\bibfnamefont {A.~I.}\ \bibnamefont {Dudenko}},
  \ and\ \bibinfo {author} {\bibfnamefont {D.}~\bibnamefont {Petrini}},\ }\href
  {\doibase 10.1088/0953-4075/35/2/315} {\bibfield  {journal} {\bibinfo
  {journal} {Journal of Physics B: Atomic, Molecular and Optical Physics}\
  }\textbf {\bibinfo {volume} {35}},\ \bibinfo {pages} {395} (\bibinfo {year}
  {2002})}\BibitemShut {NoStop}%
\bibitem [{\citenamefont {Kochur}\ and\ \citenamefont
  {Popov}(2006)}]{AGKochur_2005}%
  \BibitemOpen
  \bibfield  {author} {\bibinfo {author} {\bibfnamefont {A.~G.}\ \bibnamefont
  {Kochur}}\ and\ \bibinfo {author} {\bibfnamefont {V.~A.}\ \bibnamefont
  {Popov}},\ }\href {\doibase
  https://doi.org/10.1016/j.radphyschem.2005.11.007} {\bibfield  {journal}
  {\bibinfo  {journal} {Radiation Physics and Chemistry}\ }\textbf {\bibinfo
  {volume} {75}},\ \bibinfo {pages} {1525} (\bibinfo {year} {2006})},\ \bibinfo
  {note} {proceedings of the 20th International Conference on X-ray and
  Inner-Shell Processes}\BibitemShut {NoStop}%
\bibitem [{\citenamefont {Wang}\ \emph {et~al.}(2021)\citenamefont {Wang},
  \citenamefont {Huang}, \citenamefont {Kondev}, \citenamefont {Audi},\ and\
  \citenamefont {Naimi}}]{Wang21}%
  \BibitemOpen
  \bibfield  {author} {\bibinfo {author} {\bibfnamefont {M.}~\bibnamefont
  {Wang}}, \bibinfo {author} {\bibfnamefont {W.}~\bibnamefont {Huang}},
  \bibinfo {author} {\bibfnamefont {F.}~\bibnamefont {Kondev}}, \bibinfo
  {author} {\bibfnamefont {G.}~\bibnamefont {Audi}}, \ and\ \bibinfo {author}
  {\bibfnamefont {S.}~\bibnamefont {Naimi}},\ }\href {\doibase
  10.1088/1674-1137/abddaf} {\bibfield  {journal} {\bibinfo  {journal} {Chinese
  Physics C}\ }\textbf {\bibinfo {volume} {45}},\ \bibinfo {pages} {030003}
  (\bibinfo {year} {2021})}\BibitemShut {NoStop}%
\bibitem [{\citenamefont {Dong}\ and\ \citenamefont {Junde}(2014)}]{DONG20141}%
  \BibitemOpen
  \bibfield  {author} {\bibinfo {author} {\bibfnamefont {Y.}~\bibnamefont
  {Dong}}\ and\ \bibinfo {author} {\bibfnamefont {H.}~\bibnamefont {Junde}},\
  }\href {\doibase 10.1016/j.nds.2014.09.001} {\bibfield  {journal} {\bibinfo
  {journal} {Nuclear Data Sheets}\ }\textbf {\bibinfo {volume} {121}},\
  \bibinfo {pages} {1} (\bibinfo {year} {2014})}\BibitemShut {NoStop}%
\bibitem [{\citenamefont {Kumar}\ \emph {et~al.}(2016)\citenamefont {Kumar},
  \citenamefont {Chen},\ and\ \citenamefont {Kondev}}]{KUMAR20161}%
  \BibitemOpen
  \bibfield  {author} {\bibinfo {author} {\bibfnamefont {S.}~\bibnamefont
  {Kumar}}, \bibinfo {author} {\bibfnamefont {J.}~\bibnamefont {Chen}}, \ and\
  \bibinfo {author} {\bibfnamefont {F.}~\bibnamefont {Kondev}},\ }\href
  {\doibase 10.1016/j.nds.2016.09.001} {\bibfield  {journal} {\bibinfo
  {journal} {Nuclear Data Sheets}\ }\textbf {\bibinfo {volume} {137}},\
  \bibinfo {pages} {1} (\bibinfo {year} {2016})}\BibitemShut {NoStop}%
\bibitem [{\citenamefont {Katakura}(2011)}]{KATAKURA2011495}%
  \BibitemOpen
  \bibfield  {author} {\bibinfo {author} {\bibfnamefont {J.}~\bibnamefont
  {Katakura}},\ }\href {\doibase 10.1016/j.nds.2011.02.001} {\bibfield
  {journal} {\bibinfo  {journal} {Nuclear Data Sheets}\ }\textbf {\bibinfo
  {volume} {112}},\ \bibinfo {pages} {495} (\bibinfo {year}
  {2011})}\BibitemShut {NoStop}%
\bibitem [{\citenamefont {Chen}(2017)}]{CHEN20171}%
  \BibitemOpen
  \bibfield  {author} {\bibinfo {author} {\bibfnamefont {J.}~\bibnamefont
  {Chen}},\ }\href {\doibase 10.1016/j.nds.2017.11.001} {\bibfield  {journal}
  {\bibinfo  {journal} {Nuclear Data Sheets}\ }\textbf {\bibinfo {volume}
  {146}},\ \bibinfo {pages} {1} (\bibinfo {year} {2017})}\BibitemShut {NoStop}%
\bibitem [{\citenamefont {Fretwell}\ \emph {et~al.}(2020)\citenamefont
  {Fretwell}, \citenamefont {Leach}, \citenamefont {Bray}, \citenamefont {Kim},
  \citenamefont {Dilling}, \citenamefont {Lennarz}, \citenamefont {Mougeot},
  \citenamefont {Ponce}, \citenamefont {Ruiz}, \citenamefont {Stackhouse},\
  and\ \citenamefont {Friedrich}}]{BeEST2020}%
  \BibitemOpen
  \bibfield  {author} {\bibinfo {author} {\bibfnamefont {S.}~\bibnamefont
  {Fretwell}}, \bibinfo {author} {\bibfnamefont {K.~G.}\ \bibnamefont {Leach}},
  \bibinfo {author} {\bibfnamefont {C.}~\bibnamefont {Bray}}, \bibinfo {author}
  {\bibfnamefont {G.~B.}\ \bibnamefont {Kim}}, \bibinfo {author} {\bibfnamefont
  {J.}~\bibnamefont {Dilling}}, \bibinfo {author} {\bibfnamefont
  {A.}~\bibnamefont {Lennarz}}, \bibinfo {author} {\bibfnamefont
  {X.}~\bibnamefont {Mougeot}}, \bibinfo {author} {\bibfnamefont
  {F.}~\bibnamefont {Ponce}}, \bibinfo {author} {\bibfnamefont
  {C.}~\bibnamefont {Ruiz}}, \bibinfo {author} {\bibfnamefont {J.}~\bibnamefont
  {Stackhouse}}, \ and\ \bibinfo {author} {\bibfnamefont {S.}~\bibnamefont
  {Friedrich}},\ }\href {\doibase 10.1103/PhysRevLett.125.032701} {\bibfield
  {journal} {\bibinfo  {journal} {Phys. Rev. Lett.}\ }\textbf {\bibinfo
  {volume} {125}},\ \bibinfo {pages} {032701} (\bibinfo {year}
  {2020})}\BibitemShut {NoStop}%
\bibitem [{\citenamefont {Voytas}\ \emph {et~al.}(2001)\citenamefont {Voytas},
  \citenamefont {Ternovan}, \citenamefont {Galeazzi}, \citenamefont {McCammon},
  \citenamefont {Kolata}, \citenamefont {Santi}, \citenamefont {Peterson},
  \citenamefont {Guimar\~aes}, \citenamefont {Becchetti}, \citenamefont {Lee},
  \citenamefont {O'Donnell}, \citenamefont {Roberts},\ and\ \citenamefont
  {Shaheen}}]{PhysRevLett.88.012501}%
  \BibitemOpen
  \bibfield  {author} {\bibinfo {author} {\bibfnamefont {P.~A.}\ \bibnamefont
  {Voytas}}, \bibinfo {author} {\bibfnamefont {C.}~\bibnamefont {Ternovan}},
  \bibinfo {author} {\bibfnamefont {M.}~\bibnamefont {Galeazzi}}, \bibinfo
  {author} {\bibfnamefont {D.}~\bibnamefont {McCammon}}, \bibinfo {author}
  {\bibfnamefont {J.~J.}\ \bibnamefont {Kolata}}, \bibinfo {author}
  {\bibfnamefont {P.}~\bibnamefont {Santi}}, \bibinfo {author} {\bibfnamefont
  {D.}~\bibnamefont {Peterson}}, \bibinfo {author} {\bibfnamefont
  {V.}~\bibnamefont {Guimar\~aes}}, \bibinfo {author} {\bibfnamefont {F.~D.}\
  \bibnamefont {Becchetti}}, \bibinfo {author} {\bibfnamefont {M.~Y.}\
  \bibnamefont {Lee}}, \bibinfo {author} {\bibfnamefont {T.~W.}\ \bibnamefont
  {O'Donnell}}, \bibinfo {author} {\bibfnamefont {D.~A.}\ \bibnamefont
  {Roberts}}, \ and\ \bibinfo {author} {\bibfnamefont {S.}~\bibnamefont
  {Shaheen}},\ }\href {\doibase 10.1103/PhysRevLett.88.012501} {\bibfield
  {journal} {\bibinfo  {journal} {Phys. Rev. Lett.}\ }\textbf {\bibinfo
  {volume} {88}},\ \bibinfo {pages} {012501} (\bibinfo {year}
  {2001})}\BibitemShut {NoStop}%
\bibitem [{\citenamefont {Ray}\ \emph {et~al.}(2002)\citenamefont {Ray},
  \citenamefont {Das}, \citenamefont {Saha}, \citenamefont {Das},\ and\
  \citenamefont {Mookerjee}}]{TaMedium}%
  \BibitemOpen
  \bibfield  {author} {\bibinfo {author} {\bibfnamefont {A.}~\bibnamefont
  {Ray}}, \bibinfo {author} {\bibfnamefont {P.}~\bibnamefont {Das}}, \bibinfo
  {author} {\bibfnamefont {S.~K.}\ \bibnamefont {Saha}}, \bibinfo {author}
  {\bibfnamefont {S.~K.}\ \bibnamefont {Das}}, \ and\ \bibinfo {author}
  {\bibfnamefont {A.}~\bibnamefont {Mookerjee}},\ }\href {\doibase
  10.1103/PhysRevC.66.012501} {\bibfield  {journal} {\bibinfo  {journal} {Phys.
  Rev. C}\ }\textbf {\bibinfo {volume} {66}},\ \bibinfo {pages} {012501}
  (\bibinfo {year} {2002})}\BibitemShut {NoStop}%
\bibitem [{\citenamefont {Moler}\ and\ \citenamefont {Fink}(1963)}]{Moler63}%
  \BibitemOpen
  \bibfield  {author} {\bibinfo {author} {\bibfnamefont {R.~B.}\ \bibnamefont
  {Moler}}\ and\ \bibinfo {author} {\bibfnamefont {R.~W.}\ \bibnamefont
  {Fink}},\ }\href {\doibase 10.1103/PhysRev.131.821} {\bibfield  {journal}
  {\bibinfo  {journal} {Phys. Rev.}\ }\textbf {\bibinfo {volume} {131}},\
  \bibinfo {pages} {821} (\bibinfo {year} {1963})}\BibitemShut {NoStop}%
\bibitem [{\citenamefont {Manduchi}\ and\ \citenamefont
  {Zannoni}(1963)}]{Manduchi63}%
  \BibitemOpen
  \bibfield  {author} {\bibinfo {author} {\bibfnamefont {G.}~\bibnamefont
  {Manduchi}}\ and\ \bibinfo {author} {\bibfnamefont {G.}~\bibnamefont
  {Zannoni}},\ }\href {\doibase 10.1007/BF02812618} {\bibfield  {journal}
  {\bibinfo  {journal} {Il Nuovo Cimento}\ }\textbf {\bibinfo {volume} {27}},\
  \bibinfo {pages} {251} (\bibinfo {year} {1963})}\BibitemShut {NoStop}%
\bibitem [{\citenamefont {B{\'e}}\ \emph {et~al.}(2004)\citenamefont {B{\'e}},
  \citenamefont {Chist{\'e}}, \citenamefont {Dulieu}, \citenamefont {Browne},
  \citenamefont {Chechev}, \citenamefont {Kuzmenko}, \citenamefont {Helmer},
  \citenamefont {Nichols}, \citenamefont {Sch{\"o}nfeld},\ and\ \citenamefont
  {Dersch}}]{DDEP_54Mn}%
  \BibitemOpen
  \bibfield  {author} {\bibinfo {author} {\bibfnamefont {M.-M.}\ \bibnamefont
  {B{\'e}}}, \bibinfo {author} {\bibfnamefont {V.}~\bibnamefont {Chist{\'e}}},
  \bibinfo {author} {\bibfnamefont {C.}~\bibnamefont {Dulieu}}, \bibinfo
  {author} {\bibfnamefont {E.}~\bibnamefont {Browne}}, \bibinfo {author}
  {\bibfnamefont {V.}~\bibnamefont {Chechev}}, \bibinfo {author} {\bibfnamefont
  {N.}~\bibnamefont {Kuzmenko}}, \bibinfo {author} {\bibfnamefont
  {R.}~\bibnamefont {Helmer}}, \bibinfo {author} {\bibfnamefont
  {A.}~\bibnamefont {Nichols}}, \bibinfo {author} {\bibfnamefont
  {E.}~\bibnamefont {Sch{\"o}nfeld}}, \ and\ \bibinfo {author} {\bibfnamefont
  {R.}~\bibnamefont {Dersch}},\ }\href
  {http://www.lnhb.fr/nuclides/Mn-54_com.pdf} {\emph {\bibinfo {title} {Table
  of Radionuclides}}},\ \bibinfo {series} {Monographie BIPM-5}, Vol.~\bibinfo
  {volume} {1}\ (\bibinfo  {publisher} {Bureau International des Poids et
  Mesures},\ \bibinfo {address} {Pavillon de Breteuil, F-92310 S{\`e}vres,
  France},\ \bibinfo {year} {2004})\ \bibinfo {note} {; {E}valuation update in
  2014}\BibitemShut {NoStop}%
\bibitem [{\citenamefont {Bambynek}(1984)}]{Bamb84}%
  \BibitemOpen
  \bibfield  {author} {\bibinfo {author} {\bibfnamefont {W.}~\bibnamefont
  {Bambynek}},\ }in\ \href@noop {} {\emph {\bibinfo {booktitle} {X-84 Proc.
  X-ray and Inner-Shell Processes in Atoms, Molecules and Solids, Leipzig Aug.
  20-23, 1984}}},\ \bibinfo {editor} {edited by\ \bibinfo {editor}
  {\bibfnamefont {A.}~\bibnamefont {Meisel}}}\ (\bibinfo  {publisher} {VEB
  Druckerei},\ \bibinfo {address} {Thomas M{\"u}nzer, Langensalza, Germany},\
  \bibinfo {year} {1984})\ \bibinfo {note} {post-deadline paper
  P-1}\BibitemShut {NoStop}%
\bibitem [{\citenamefont {Bambynek}(1967)}]{Bamb67om}%
  \BibitemOpen
  \bibfield  {author} {\bibinfo {author} {\bibfnamefont {W.}~\bibnamefont
  {Bambynek}},\ }\href {\doibase 10.1007/BF01325886} {\bibfield  {journal}
  {\bibinfo  {journal} {Zeitschrift f{\"u}r Physik}\ }\textbf {\bibinfo
  {volume} {206}},\ \bibinfo {pages} {66} (\bibinfo {year} {1967})}\BibitemShut
  {NoStop}%
\bibitem [{\citenamefont {Pengra}\ \emph {et~al.}(1972)\citenamefont {Pengra},
  \citenamefont {Genz}, \citenamefont {Renier},\ and\ \citenamefont
  {Fink}}]{Pengra72}%
  \BibitemOpen
  \bibfield  {author} {\bibinfo {author} {\bibfnamefont {J.~G.}\ \bibnamefont
  {Pengra}}, \bibinfo {author} {\bibfnamefont {H.}~\bibnamefont {Genz}},
  \bibinfo {author} {\bibfnamefont {J.~P.}\ \bibnamefont {Renier}}, \ and\
  \bibinfo {author} {\bibfnamefont {R.~W.}\ \bibnamefont {Fink}},\ }\href
  {\doibase 10.1103/PhysRevC.5.2007} {\bibfield  {journal} {\bibinfo  {journal}
  {Phys. Rev. C}\ }\textbf {\bibinfo {volume} {5}},\ \bibinfo {pages} {2007}
  (\bibinfo {year} {1972})}\BibitemShut {NoStop}%
\bibitem [{\citenamefont {Loidl}\ \emph {et~al.}(2018)\citenamefont {Loidl},
  \citenamefont {Rodrigues},\ and\ \citenamefont {Mariam}}]{Loidl18}%
  \BibitemOpen
  \bibfield  {author} {\bibinfo {author} {\bibfnamefont {M.}~\bibnamefont
  {Loidl}}, \bibinfo {author} {\bibfnamefont {M.}~\bibnamefont {Rodrigues}}, \
  and\ \bibinfo {author} {\bibfnamefont {R.}~\bibnamefont {Mariam}},\ }\href
  {\doibase 10.1016/j.apradiso.2017.10.042} {\bibfield  {journal} {\bibinfo
  {journal} {Applied Radiation and Isotopes}\ }\textbf {\bibinfo {volume}
  {134}},\ \bibinfo {pages} {395} (\bibinfo {year} {2018})}\BibitemShut
  {NoStop}%
\bibitem [{\citenamefont {B{\'e}}\ \emph {et~al.}(2016)\citenamefont {B{\'e}},
  \citenamefont {Chist{\'e}}, \citenamefont {Dulieu}, \citenamefont {Kellett},
  \citenamefont {Mougeot}, \citenamefont {Arinc}, \citenamefont {Chechev},
  \citenamefont {Kuzmenko}, \citenamefont {Kib{\'e}di}, \citenamefont {Luca},\
  and\ \citenamefont {Nichols}}]{DDEP_109Cd}%
  \BibitemOpen
  \bibfield  {author} {\bibinfo {author} {\bibfnamefont {M.-M.}\ \bibnamefont
  {B{\'e}}}, \bibinfo {author} {\bibfnamefont {V.}~\bibnamefont {Chist{\'e}}},
  \bibinfo {author} {\bibfnamefont {C.}~\bibnamefont {Dulieu}}, \bibinfo
  {author} {\bibfnamefont {M.}~\bibnamefont {Kellett}}, \bibinfo {author}
  {\bibfnamefont {X.}~\bibnamefont {Mougeot}}, \bibinfo {author} {\bibfnamefont
  {A.}~\bibnamefont {Arinc}}, \bibinfo {author} {\bibfnamefont
  {V.}~\bibnamefont {Chechev}}, \bibinfo {author} {\bibfnamefont
  {N.}~\bibnamefont {Kuzmenko}}, \bibinfo {author} {\bibfnamefont
  {T.}~\bibnamefont {Kib{\'e}di}}, \bibinfo {author} {\bibfnamefont
  {A.}~\bibnamefont {Luca}}, \ and\ \bibinfo {author} {\bibfnamefont
  {A.}~\bibnamefont {Nichols}},\ }\href
  {http://www.lnhb.fr/nuclides/Cd-109_com.pdf} {\emph {\bibinfo {title} {Table
  of Radionuclides}}},\ \bibinfo {series} {Monographie BIPM-5}, Vol.~\bibinfo
  {volume} {8}\ (\bibinfo  {publisher} {Bureau International des Poids et
  Mesures},\ \bibinfo {address} {Pavillon de Breteuil, F-92310 S{\`e}vres,
  France},\ \bibinfo {year} {2016})\BibitemShut {NoStop}%
\bibitem [{\citenamefont {B{\'e}}\ \emph {et~al.}(2011)\citenamefont {B{\'e}},
  \citenamefont {Chist{\'e}}, \citenamefont {Dulieu}, \citenamefont {Mougeot},
  \citenamefont {Chechev}, \citenamefont {Kuzmenko}, \citenamefont {Kondev},
  \citenamefont {Luca}, \citenamefont {Gal{\'a}n}, \citenamefont {Nichols},
  \citenamefont {Arinc}, \citenamefont {Pearce}, \citenamefont {Huang},\ and\
  \citenamefont {Wang}}]{DDEP_125I}%
  \BibitemOpen
  \bibfield  {author} {\bibinfo {author} {\bibfnamefont {M.-M.}\ \bibnamefont
  {B{\'e}}}, \bibinfo {author} {\bibfnamefont {V.}~\bibnamefont {Chist{\'e}}},
  \bibinfo {author} {\bibfnamefont {C.}~\bibnamefont {Dulieu}}, \bibinfo
  {author} {\bibfnamefont {X.}~\bibnamefont {Mougeot}}, \bibinfo {author}
  {\bibfnamefont {V.}~\bibnamefont {Chechev}}, \bibinfo {author} {\bibfnamefont
  {N.}~\bibnamefont {Kuzmenko}}, \bibinfo {author} {\bibfnamefont
  {F.}~\bibnamefont {Kondev}}, \bibinfo {author} {\bibfnamefont
  {A.}~\bibnamefont {Luca}}, \bibinfo {author} {\bibfnamefont {M.}~\bibnamefont
  {Gal{\'a}n}}, \bibinfo {author} {\bibfnamefont {A.}~\bibnamefont {Nichols}},
  \bibinfo {author} {\bibfnamefont {A.}~\bibnamefont {Arinc}}, \bibinfo
  {author} {\bibfnamefont {A.}~\bibnamefont {Pearce}}, \bibinfo {author}
  {\bibfnamefont {X.}~\bibnamefont {Huang}}, \ and\ \bibinfo {author}
  {\bibfnamefont {B.}~\bibnamefont {Wang}},\ }\href
  {http://www.lnhb.fr/nuclides/I-125_com.pdf} {\emph {\bibinfo {title} {Table
  of Radionuclides}}},\ \bibinfo {series} {Monographie BIPM-5}, Vol.~\bibinfo
  {volume} {6}\ (\bibinfo  {publisher} {Bureau International des Poids et
  Mesures},\ \bibinfo {address} {Pavillon de Breteuil, F-92310 S\`evres,
  France},\ \bibinfo {year} {2011})\BibitemShut {NoStop}%
\bibitem [{\citenamefont {Leutz}\ and\ \citenamefont
  {Ziegler}(1964)}]{Leutz64}%
  \BibitemOpen
  \bibfield  {author} {\bibinfo {author} {\bibfnamefont {H.}~\bibnamefont
  {Leutz}}\ and\ \bibinfo {author} {\bibfnamefont {K.}~\bibnamefont
  {Ziegler}},\ }\href {\doibase 10.1016/0029-5582(64)90237-8} {\bibfield
  {journal} {\bibinfo  {journal} {Nuclear Physics}\ }\textbf {\bibinfo {volume}
  {50}},\ \bibinfo {pages} {648} (\bibinfo {year} {1964})}\BibitemShut
  {NoStop}%
\bibitem [{\citenamefont {Quarati}\ \emph {et~al.}(2016)\citenamefont
  {Quarati}, \citenamefont {Dorenbos},\ and\ \citenamefont
  {Mougeot}}]{Quarati16}%
  \BibitemOpen
  \bibfield  {author} {\bibinfo {author} {\bibfnamefont {F.}~\bibnamefont
  {Quarati}}, \bibinfo {author} {\bibfnamefont {P.}~\bibnamefont {Dorenbos}}, \
  and\ \bibinfo {author} {\bibfnamefont {X.}~\bibnamefont {Mougeot}},\ }\href
  {\doibase 10.1016/j.apradiso.2016.01.017} {\bibfield  {journal} {\bibinfo
  {journal} {Applied Radiation and Isotopes}\ }\textbf {\bibinfo {volume}
  {109}},\ \bibinfo {pages} {172} (\bibinfo {year} {2016})}\BibitemShut
  {NoStop}%
\end{thebibliography}%
	
\end{document}